\documentclass{article}
\usepackage[utf8]{inputenc}
\usepackage{graphicx}
\usepackage{algpseudocode}
\usepackage{algorithm}
\usepackage{amsmath}
\usepackage{amssymb}
\usepackage{amsthm}
\usepackage{caption}
\usepackage{subcaption}
\usepackage{mathtools}
\usepackage[dvipsnames]{xcolor}
\usepackage{adjustbox}
\usepackage[margin=1in]{geometry}

\usepackage[draft]{todonotes}

\definecolor{Lavender}{RGB}{179, 169, 235}
\newcommand{\bluesq}{\raisebox{.17em}{\text{\fcolorbox{Cerulean}{Cerulean}{\rule{0pt}{2pt}\rule{2pt}{0pt}}}}}
\newcommand{\greensq}{\raisebox{.17em}{\text{\fcolorbox{LimeGreen}{LimeGreen}{\rule{0pt}{2pt}\rule{2pt}{0pt}}}}}
\newcommand{\orangesq}{\raisebox{.17em}{\text{\fcolorbox{YellowOrange}{YellowOrange}{\rule{0pt}{2pt}\rule{2pt}{0pt}}}}}
\newcommand{\yellowsq}{\raisebox{.17em}{\text{\fcolorbox{Goldenrod}{Goldenrod}{\rule{0pt}{2pt}\rule{2pt}{0pt}}}}}
\newcommand{\lavsq}{\raisebox{.17em}{\text{\fcolorbox{Lavender}{Lavender}{\rule{0pt}{2pt}\rule{2pt}{0pt}}}}}
\newcommand{\death}{\raisebox{-0.17em}{\includegraphics[width=3.3mm]{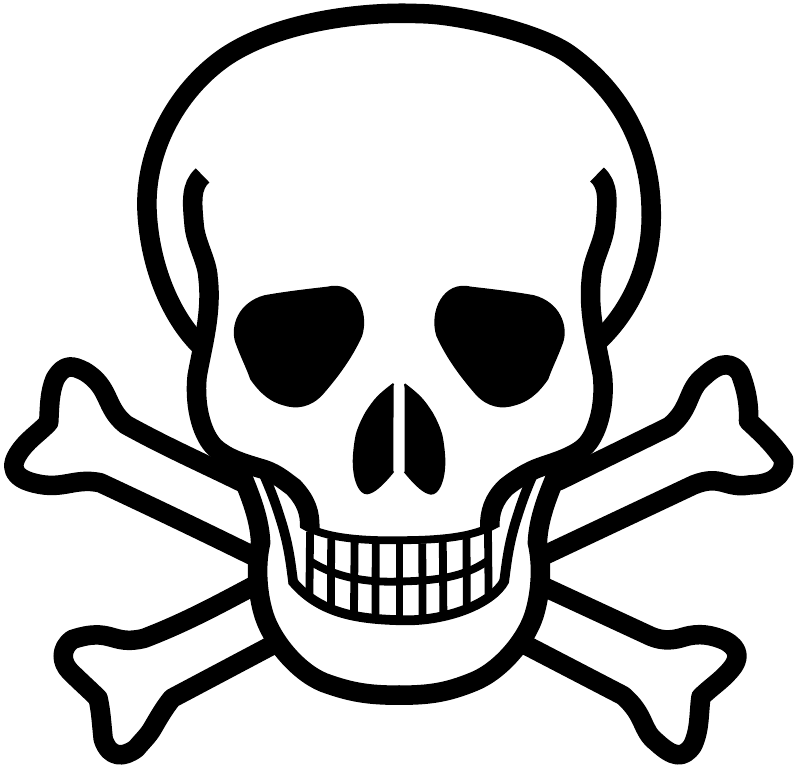}}}

\title{Multiscale Parallel Tempering for Fast Sampling on Redistricting Plans}
\author{Gabriel Chuang, Gregory Herschlag, Jonathan Mattingly}

\begin{document}
\maketitle 
\begin{abstract}
When auditing a redistricting plan, a persuasive method is to compare the plan with an ensemble of neutrally drawn redistricting plans. Ensembles are generated via algorithms that sample distributions on balanced graph partitions. To audit the partisan difference between the ensemble and a given plan, one must ensure that the non-partisan criteria are matched so that we may conclude that partisan differences come from bias rather than, for example, levels of compactness or differences in community preservation. Certain sampling algorithms allow one to explicitly state the policy-based probability distribution on plans, however, these algorithms have shown poor mixing times for large graphs (i.e. redistricting spaces) for all but a few specialized measures. In this work, we generate a multiscale parallel tempering approach that makes local moves at each scale. The local moves allow us to adopt a wide variety of policy-based measures. We examine our method in the state of Connecticut and succeed at achieving fast mixing on a policy-based distribution that has never before been sampled at this scale. Our algorithm shows promise to expand to a significantly wider class of measures that will (i) allow for more principled and situation-based comparisons and (ii) probe for the typical partisan impact that policy can have on redistricting.
\end{abstract}

\section{Introduction}
Over the past decade, sampling techniques have become a critical tool in auditing redistricting plans for partisan and racial gerrymandering, both as an academic pursuit
\cite{ChenRodden13,Liu16,Chikina_Frieze_Pegden_2017,chikina2019separating,herschlag2020quantifying,fifield2020automated,moonVa,deford2019recombination,autry2023metropolized,autry2021metropolized,mccartan2020sequential,gurnee2021fairmandering,cannon2022spanning} and in court cases \cite{RuchoVCC,LWVvPA,LewisVCommonCause,HarperVHallMoore,AllenVMilligan}. The goal of such audits is to compare the partisan behavior or racial makeup of a given plan to an ensemble of (typically neutrally-drawn) plans; if the plan in question differs from the ensemble in a significant way, the given plan might be labeled extreme or at a minimum the map maker might be asked to defend the differences.

However, when comparing an enacted plan to an ensemble, one must ensure that the plans generated by a sampling algorithm reflect the same policy goals as the plan to be audited. For example, if all the plans in the ensemble are significantly more compact (by some definition) than the plan to be audited, then it is not clear whether any partisan differences between the ensemble and the plan are due to mapmaker intent, rather than some intrinsic relationship between compactness and partisan outcomes. In short, one must align non-partisan criteria between maps in order to audit partisan bias. %\jcm{Maybe mention that simply restricting to maps with in a set is not sufficient to match the plan or policy under consideration} -- also mentioned this now explicitly with a word of caution in the sampling problem section

The fundamental question in auditing a given plan using this methodology is ``Would a random map with similar (non-partisan) \emph{policy} considerations typically have similar partisan (or racial) qualities to the map in question?'' Randomness assumes a probability distribution on redistricting plans that is informed by the expressed policy considerations in a given map. We therefore stress that the ensembles are a tool to analyze the consequences of a policy. We must choose methodologies of generating ensembles in a way that is consistent with the stated and expressed (non-partisan) policy considerations of map makers in order to make appropriate comparisons between an ensemble and an enacted plan. 

Fundamentally, ensembles and sampling methods arise via algorithms that encode policy (via a probability distribution) and serve as a method to access the encoded policy via Monte Carlo methods. %For example, some algorithms will produce plans that are more compact than other % the prior sentence is very opaque to me -gabriel; better now? gjh
For any given sampling methodology, there is a probability distribution on the space of redistricting plans that it samples from. For some samplers, the distribution is implicit \cite{ChenRodden13,Liu16,herschlag2020quantifying,deford2019recombination,gurnee2021fairmandering}, whereas for others it is explicit \cite{MattinglyVaughn2014,Chikina_Frieze_Pegden_2017,chikina2019separating,fifield2020automated,autry2023metropolized,autry2021metropolized,mccartan2020sequential,cannon2022spanning}. Some of the key elements in these samplers involve various Monte Carlo methods such as the construction of Markov Chain methods via Metropolis-Hastings based on flipping boundary nodes (also called single node flip or flip walk; e.g. see \cite{MattinglyVaughn2014,Chikina_Frieze_Pegden_2017,chikina2019separating,fifield2020automated,herschlag2020quantifying,deford2019recombination,najt2021empirical}). More recently there has been significant interest and attention in using tree-based methods that draw and cut spanning trees to rearrange redistricting plans.  This process has been developed via Markov Chain Monte Carlo methods \cite{deford2019recombination,autry2023metropolized,autry2020multiscale,cannon2022spanning} and via Sequential Monte Carlo Methods \cite{mccartan2020sequential}. 
Sampling redistricting plans is formally cast as a balanced graph partitioning problem, in which the nodes represent some essential administrative unit like a census block or precinct, and edges represent adjacency between units. 

The boundary node flip MCMC methods have proven successful on small graphs but have failed to mix as the graph size grows (see e.g. \cite{jcmReport,njatDedfordSolomon2019graphs,najt2021empirical}). The tree-based measures have shown promising mixing rates on large graphs, but only when sampling from measures that prefer plans with higher numbers of associated spanning forests \cite{deford2019recombination,autry2023metropolized,autry2021metropolized,mccartan2020sequential}. 
%
% \jcm{Such choices can have implications beyond compactness. See  undergraduate report} -- did this below when introducing the measure/sampling problem
%
Although such measures define one possible type of policy-based probability distribution, there are many other distributions of interest that do not focus on the number of associated spanning forests. Furthermore, some redistricting commissions explicitly state the policies they are sampling (e.g. that they should use the traditional Polsby-Popper and Reock scores for compactness) which have been shown to be only weakly correlated with spanning forest counts \cite{autry2023metropolized}. There has been work using the tree-based proposal moves with parallel tempering to refocus the space on plans with tighter Polsby-Popper scores \cite{zhao2022mathematically}; however, these works maintain the focus on partitions with high forests as a base measure on the space.

We take the position that it is of upmost importance to draw plans from a known, policy-driven distribution that can be modified based on the policy expression of a redistricting body. For example, one may specify a preference for plans that have a particular degree of compactness and preserve communities of interest. If we can maintain efficient sampling over a wider class of distributions, then we can (i) assess the impact of various policy considerations and (ii) ensure that our distributions reflect the stated policies of a redistricting body and expression of these policies as seen in an audited plan.

% There has also been extensive work using multi-scale methods to embed ideas of preserving areas of interest. More broadly, multiscale methods have been successful at [... what?]. 

In the current state of the field, we already have access to a large array of policies for small graphs. It has been consistently observed both for the single node boundary walk methods and the tree based methods that they can efficiently sample various policies on small graphs (for an example on the node flip see \cite{jcmReport}; for an example on the tree based method see \cite{jcmReportHarperVHallMoore}). This together with the recent work on multiscale sampling \cite{autry2021metropolized} leads to the following question: Can we somehow coarsen the graph to access the fast mixing properties of the existing algorithms, and then refine it back down to the full state space?

In this paper, we begin to answer this question in the affirmative by introducing a multi-scale sampler that employs local moves (via single node flips at district boundaries) on a hierarchy of coarsened graphs. 
% A major advantage of single node flip is that it is a local move that is adapable to a wide array of measures. 
This method allows us to use boundary flip methods at very coarse levels to mix, and transmit this mixing down to finer scale levels of the hierarchy by linking the levels via parallel tempering. 

Traditionally, tempering schemes are employed on identical state spaces; however, our coarsened hierarchy samples plans from different, and potentially disjoint, state spaces. Finer scales may split coarsened nodes (that must be kept whole at the coarser level); at coarser scales we relax the population balance constraints so that the coarser nodes with larger populations may adequately fluctuate to promote mixing. Therefore, when proposing to swap states between a fine and coarse level, we develop a paired method to jointly project the state of each chain onto the state space of the other in a tractable and probabilistic way. 

We demonstrate our sampler on the five congressional districts of Connecticut. We use target measures that considers contiguity, population balance, and compactness defined by the Polsby-Popper score. We sample the congressional districts over the roughly 700 precincts in Connecticut and remark that accessing this class of measure at this scale has never before been achieved. 

We also demonstrate that our chains converge efficiently and further demonstrate that we have the ability to control the degree to which we prioritize Polsby-Popper compactness. 
We also compare our result to sampling over the space weighted by the number of spanning forests; we show that there is significant discrepancy between the forest-based sampler and the new measures, both in terms of typical partisan outcomes of the ensembles and in terms of how the measures prioritize compactness.
% We also compare our chains to the chain run at the measure that is weighted by the number of associated spanning forests; we show that (i) the levels of observed Polsby-Popper compactness scores are disjoint and (ii) that there is a large discrepancy between the typical range in the observed partisan properties of the maps under historic elections.  
% In performing this comparison, we demonstrate that the policy-based probability distribution that one chooses can have a significant impact both on the non-partisan qualities of the ensemble, and also on the partisan qualities of the ensemble. 
% effect means that plans that one would consider to be an outlier in terms of partisan performance.\jcm{But if you had used reversible recom you could have tuned it to sample that PP compactness?}

In short, our sampler mixes efficiently on a relevant class of measures at an unprecedented scale. Because of this, it (i) allows us to contrast the impact of different policy considerations and (ii) significantly expands our ability to audit redistricting plans.

% Discussion of Recom/SNF; Discussion of success of multi-scale methods; introduce our method: ideas and successes

\section{Overview}

%Broad high-level description of our method.  1-2 pages of what you would want someone being introduced to the work to understand. Include pretty pedagogical figures

%e.g. Want to sample a precinct graph, create an arbitrary hierarchy; create levels which slowly merge elements in the hierarchy, use parallel tempering to swap levels; Very brief (with cartoon) develop method of probablistically exchanging population with hierarchy preservation.

In redistricting, the task is to partition a set of atomic units---such as precincts, census blocks, or counties---into a fixed number of \emph{districts}, each of which must be connected\footnote{via Queen, Rook, or legally defined adjacency, depending on the state} and contain the same population. Various other properties may be desirable or legally mandated, depending on the state, such as creating compact districts, creating a specified number of majority-minority districts (or districts that comply with the Voting Rights Act), and creating districts that preserve counties and/or communities. Together, the stated desires or mandates in redistricting form a redistricting \emph{policy}.

The policy may be quantified by formulating a score function that encodes a relative preference between two plans, e.g. ``Plan A is 10 times `more desirable' than Plan B.'' This quantified preference may be thought of as an un-normalized probability distribution across redistricting plans. When auditing a given district plan, or when seeking to understand the implications of a given quantified policy, one may utilize Monte Carlo methods to understand the space of redistricting plans that comply with or are centered around the policy-based probability distribution.

%MOVE THIS TO INTRODUCTION Generating Monte Carlo samples for generic policies is a major open problem in the field of redistricting. The two most commonly mentioned methods are the use of Single Node Flips (also called Flip Walks and Boundary Diffusion methods) \cite{} and Recombination along with variants \cite{}. The former has been shown to face barriers in sampling (in particular, slow mixing on large graphs due to energetic barriers in the policies) \cite{} % non-rev paper whereas the later naturally targets distributions on spanning forests rather than partitions and has thus only been demonstrated to generate samples successfully when the policy is aligned with sampling from balanced spanning forests.

\begin{figure}[ht]
    \centering
    \begin{subfigure}[b]{0.185\textwidth}
        \includegraphics[width=0.75\textwidth]{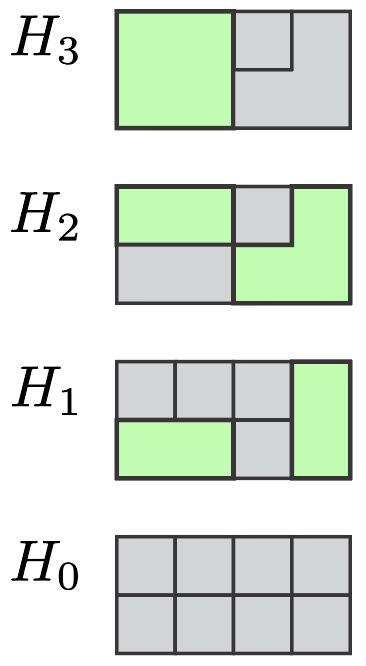}
        \caption{4-level hierarchy on a 2x4 grid graph. Merged nodes at each level are highlighted in green.}
        \label{fig:hierarchyExplainer}
    \end{subfigure}
    \hfill 
    \begin{subfigure}[b]{0.8\textwidth}
        \includegraphics[width=\textwidth]{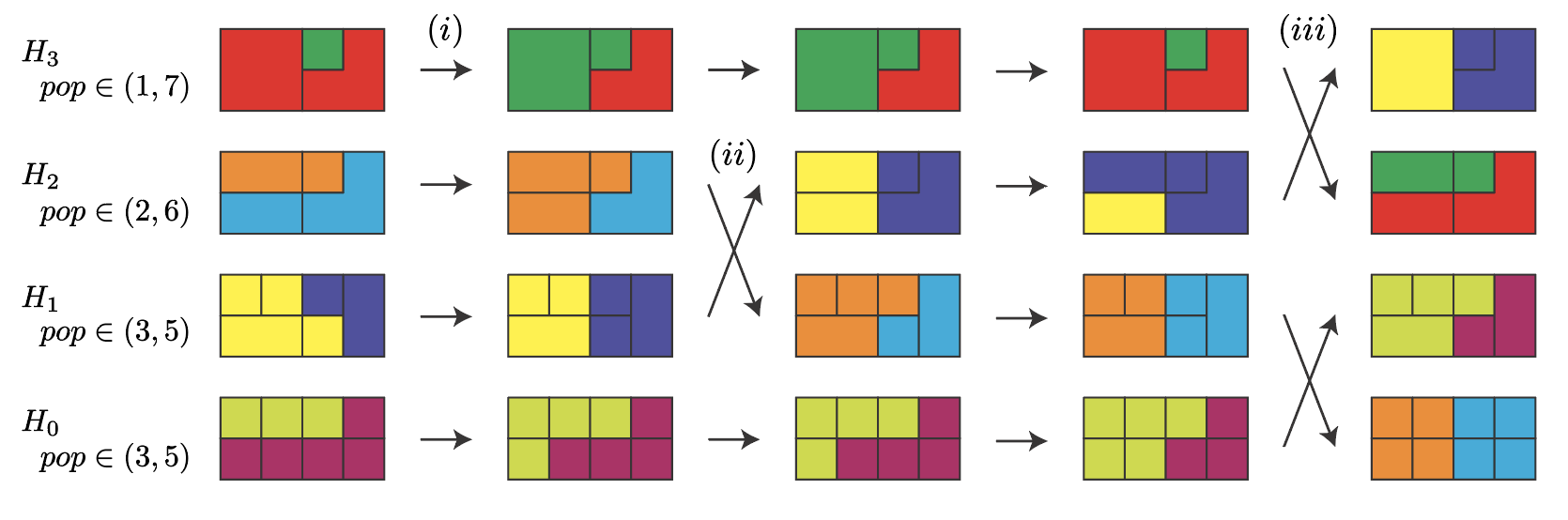}
        \caption{High-level sketch of the overall procedure. A single-node-flip Markov Chain is run at each level of the hierarchy, and partitions are swapped between adjacent levels of the hierarchy at fixed intervals. The target measure is placed on the finest (bottom-most) level (in this case, population bounds of $(3,5)$), and the measure is relaxed towards the coarser levels of the hierarchy.} 
        \label{fig:ptExplainer}
    \end{subfigure}
    \caption{Example hierarchy and overall method for 2 districts on a 2x4 grid graph. (For this example, each unit square has population 1.)}
    \label{fig:explainer}
\end{figure}

In this work, we introduce a multi-scale parallel tempering algorithm and demonstrate numerical evidence that it is capable of sampling from an expanded class of policy-based probability distributions, compared with with the existing literature. The multi-scale framework is based on the generation of a hierarchy. % i don't like the prior sentence (it's not based on the generation of a hierarchy), but am struggling to make it better - gabriel
The bottom level of the hierarchy is the original atomic graph that we wish to sample from, e.g. the precinct graph of the state of interest. Each subsequent level is formed by merging a subset of the nodes in the previous level. By gradually merging nodes as we travel up the hierarchy, we eventually arrive at a coarse contraction of the original graph, with a much smaller node count than the original graph (e.g. see Figure~\ref{fig:hierarchyExplainer}). The hierarchy, in principle, may be constructed arbitrarily; however, we empirically discover and report several beneficial heuristics.

We place a probability measure on each level of the hierarchy. The bottom level's measure is the measure that best encapsulates the policy and which we will use to build our ensemble of sampled maps. The subsequent measures are formulated so that (i) sampling becomes easier as the graph becomes more coarse (e.g. via looser constraints), and (ii) so the measures at adjacent levels are similar. The latter condition allows us to employ a parallel tempering algorithm so that consecutive measures have a better chance of exchanging their current states. The idea is that the better mixing and exploration of the coarser levels can be transferred to the finest level by the swapping of adjacent levels.

Each level of the hierarchy samples from a distinct space of plans. For example, plans at the finer scales may have tighter population bounds, whereas plans at coarse scales require certain fine-scale nodes to be assigned to the same district. Thus the partitions at consecutive chains are sampled from different, and potentially disjoint, state spaces. One key technical aspect of our method that enables tempering is the introduction of a swap operation that takes in partitions from a given level and probabilistically (and tractably) alters them to fit into another level's underlying state-space. Importantly, the probability of a succesful swap is chosen so that the marginal measure at each level is preserved.  In rare cases, the two partitions we hope to swap can do so with no conflicts (e.g., $(ii)$ in Figure~\ref{fig:ptExplainer}). However, since each level of the hierarchy is associated with a distinct graph, one typically cannot simply exchange a partition from one level to another. For example, a partition at level $i$ may assign two nodes to different districts when those two nodes are merged at level $i+1$ (as in the yellow-blue partition before swap $(iii)$ in Figure~\ref{fig:ptExplainer}), or a partition at level $j+1$ may have districts that fall outside the population bounds at level $j$ (as in the red-green partition before swap $(iii)$ in Figure~\ref{fig:ptExplainer}). We use a reversible paired exchange mechanism, described in Section~\ref{ssec:SwapMech} and Appendix~\ref{apdx:Swap}, to adjust partitions to be consistent with the post-swap levels/measures. 

% We place a different measure at each level of the hierarchy. The finest level samples from the target measure, and successively coarser levels sample from increasingly relaxed versions of the measure (e.g., larger allowable population bounds, lower weight on compactness, etc.). 
We incrementally vary the measures (larger allowable population bounds, lower weight on compactness, etc.) from one level to the next. This keeps the individual swap probabilities reasonable while giving the chains more freedom to make large moves at the coarse levels (e.g., $(i)$ in Figure~\ref{fig:ptExplainer}), enabling faster exploration of the space. The final ensemble is drawn from the finest level of the hierarchy (i.e. from the target measure), as is standard in the parallel tempering framework. 

% The measure that we sample from does \emph{not} depend on the choice of hierarchy. This enables us to choose a hierarchy with targetted properties (for example, we can aim to have relatively population-balanced nodes, or create a hierarchy that prioritizes respecting county boundaries). % maybe unneeded here? -gabriel 

% The swap opperator allows us to implement standard parallel tempering techniques (via Metropolis-Hastings) on distinct spaces of graph partitions. By doing so, we allow the exchange between partitions with coarse nodes between those with tighter population requirements with finer nodes. % this sentence doesn't parse for me -- gabriel 

In between swap proposals, we run Metropolized Markov Chains at each level for a fixed number of steps. In principle, we may use any Markov Chain at each level, however, in this work we exclusively employ a single-node-flip Markov chain at each level of the hierarchy. Each chain spends some time at each level of the hierarchy: sometimes making small, local moves at the finer levels, and sometimes making larger, global moves at the coarser levels. The method is depicted in Figure~\ref{fig:ptExplainer}. Single-node-flip algorithms make local moves that are relatively agnostic to the choice of measure. In this respect it is ideal as a proposal chain in a metropolization scheme to sample from a wide of measures. However, as already mentioned,  single-node-flip can suffer from slow mixing when the sytem size is large.  Our multi-scale framework allows us to overcome the energetic barriers that typically result in slow mixing for this method.

%To our knowledge, this is the first algorithm that has emperically demonstrated mixing results on large real-world redistricting graphs on measures (i.e. policies) that go beyond those that focus around spanning forests. -- maybe put this in the intro instead.

\section{The Sampling Problem}\label{sec:SamplingProblem}
% Formalize graph, measure, and the problem. 

% The setting and notation we use here is largely the same as in prior work, e.g. \cite{herschlag2020quantifying,autry2023metropolized}. 
Let $G$ be a graph with vertices $V$ and edges $E$, which we wish to partition into $d$ districts. Each vertex represents a geographic region, such as a precinct, county, or census block, that we wish to assign to a district. Each edge represents adjacency. For this work, we focus on planar graphs, although the techniques we discuss generalize to general graphs as well. 

A \emph{redistricting plan} of $G$ is a function $\xi : V \to \{1, 2, 3, \cdots d\}$. We will use $\xi^{(i)}$ to denote the subgraph consisting of the $i$th district; i.e., $V(\xi^{(i)}) = \{v \in V : \xi(v) = i\}$ and $E(\xi^{(i)}) = \{(u,v) \in E : \xi(u) = \xi(v) = i\}$. We will require that each district be contiguous, i.e., that each $\xi^{(i)}$ has exactly one connected component. When necessary, we may use $\xi_G$ and $\xi^{(i)}_G$ to specify the graph on which the redistricting plan or a district is drawn, respectively. 

% To check: may want to add hard constraints explicitly to $\Xi$
We will use $\Xi(G, d, J)$ to denote the space of valid $d$-partitions of graph $G$ subject to constraints imposed by a score function $J$. 
%When  $G$, $d$, and $J$ are clear from context, we may omit them and use simply $\Xi$. when $G$ is associated with some $J$ and $d$ is clear from context we may write $\Xi(G)$. 
Note that $\Xi(G, d, J)$ is \emph{not} all functions $\xi:V(G) \to \{1, 2, \cdots d\}$, but only those that satisfy the specified constraints which are specified with $J(\xi)<\infty$, such as contiguity, population balance, etc. When we do not wish to consider a constrained space, we simply write $\Xi(G, d)$. 

The nodes and edges of $G$ may be augmented with additional information to allow for evaluation of various redistricting criteria, such as population balance, compactness, etc. In particular, each node may have a population and an area; each edge may have a length (i.e., the border length between the two regions). These quantities are also applied at the district level; that is, 
\begin{align}
\text{pop}(\xi^{(i)}) = \sum_{\mathclap{v \in \xi^{(i)}}} \text{pop}(v),\quad \text{area}(\xi^{(i)}) = \sum_{\mathclap{v \in \xi^{(i)}}} \text{area}(v), \quad\text{and} \quad\text{length}(\xi^{(i)}, \xi^{(j)}) = \sum_{\mathclap{u \in \xi^{(i)}, v \in \xi^{(j)}}} \text{length}(u,v).
\end{align}

We use such quantities to specify the probability measures we wish to sample from. Our technique is intended to sample from a wide class of desired measures; one may in general want to place a variety of different energy functions, $J$, on redistricting plans and sample according to the probability measure 
\begin{align}
P(\xi) \propto e^{-J(\xi)}.
\end{align}

The energy $J$ is used to capture policy-based characteristics we wish the districts to have. For example, we may have a hard constraint on how far each district's population can deviate from the ideal population:
\begin{align}
J_{\text{pop}}(\xi; \delta_\text{pop}) = \begin{cases}
    0 & \frac{\text{total population}}{\# \text{districts}} - \delta_\text{pop} \leq \text{pop}(\xi_i) \leq \frac{\text{total population}}{\# \text{districts}} + \delta_\text{pop} \text{ for all }\xi_i \\ 
    \infty & \text{otherwise}
\end{cases}.
\label{eq:popconstraint}
\end{align}
We will also assume that $J$ contains a contiguity constraint (i.e. we will add $J_{contiguous}$ which is zero if the districts/partitions are contiguous and infinity if not).

Another common requirement is that districts be \emph{compact}. Compactness in two dimensional geometries has a multitude of definitions \cite{montero2009state}; The Polsby-Popper \cite{polsby1991third} and Reock \cite{reock1961note} scores are perhaps still the most commonly reported scores by redistricting bodies, and some of these bodies explicitly state their use as part of the criteria (e.g. see \cite{moncrief2011reapportionment}). The Polsby-Popper score is the reciprocal of the isoperimetric ratio scaled so that it varies between zero and one, whereas the Reock score is the ratio of area between a district and the smallest inscribing circle. More recently, some groups have begun to advocate for counting the number of cut edges in the graph, in part due to the compatibility with tree-based sampling methods \cite{deford2019recombination}. It has been observed that cut-edge metrics also enforce certain additional policies as cutting through large rural precincts tends to cut fewer edges than when district lines traverse through many more compact urban precincts \cite{Eichenlaub22}.
There has also been a recently proposed machine learning approach to identifying compactness called ``You know it when you see it'' which has been demonstrated to be consistent with human intuition of compactness \cite{kaufman2021measure}.  Furthermore, it has been demonstrated that there can be little correlation between compactness scores \cite{kaufman2021measure}. Given the abundance of ways to measure compactness, one may wish to ask what is a typical consequence of selecting from various choices.

In order to combine criteria, we adopt a weighted sum of the criteria, e.g., 
\begin{align}
J(\xi) = J_{\text{contiguous}}(\xi) + J_{\text{pop}}(\xi; \delta_\text{pop}) + w_{comp} J_\text{comp}(\xi),
\label{eqn:scorefunc}
\end{align}
where we have not added a weight to $J_{\text{contiguous}}$ or $J_{pop}$ due to these being constraints with a value of 0 or $\infty$. We can also choose a measure like $J(\xi) = J_{\text{pop}}(\xi) + \mathbf{1}_{\{J_\text{comp}(\xi)<c\}}(\xi)$ which places a hard threshold constraint on the compactness score and recovers the uniform measure on a subspace of graph partitions.\footnote{We note that one should use caution on such a measure: Due to the high demensionality of these spaces, one may always find a uniform sample of $\xi$ on this space very close to the threshold value $c$} Note that this flexibility has always been the stated goal in past work, however it has been unattainable for many definitions of $J$.

It is possible to add other types of energies, either as hard constraints or as weighted preferences. Such forms may involve various definitions of county, municipal, or community preservation, or scores that account for the voting rights act in various ways. 

In the current work, we primarily adopt the Polsby-Popper compactness score along with a hard population constraint. We define the compactness score as 
\begin{align}
J_\text{comp}^{PP}(\xi) = \sum_{i=1}^d \frac{\sum_{j\neq i} \text{length}(\xi^{(i)}, \xi^{(j)})^2}{\text{area}(\xi^{(i)})}.
\label{eq:PP}
\end{align}
The reason for this choice is that (i) Polsby-Popper scores are commonly used in practice, and (ii) sampling large graphs from such measures has not yet been achieved; we therefore demonstrate that our methods are capable of sampling from a new class of policy-based probability distributions. In \eqref{eq:PP}, we have employed the sum of the reciprocals reciprocal of the Polsby-Popper score for each district. The  reciprocals reciprocal of the Polsby-Popper score coincides with the isoperimetric constant for the district.

\section{A Parallel Tempering Multiscale Approach}

In this section we present our parallel tempering multiscale approach. The method involves (i) generating a hierarchy of coarsened graphs, (ii) developing a swap mechanism to exchange information across levels of the hierarchy, and (iii) establishing measures for each level of the hierarchy.  The swap mechanism and probabilities must be carefully constructed to preserve the desired measures at each level.
\subsection{Geographic Hierarchy} 

\begin{figure}[ht]
    \centering
    \begin{subfigure}[b]{0.48\textwidth}
        \includegraphics[width=\textwidth]{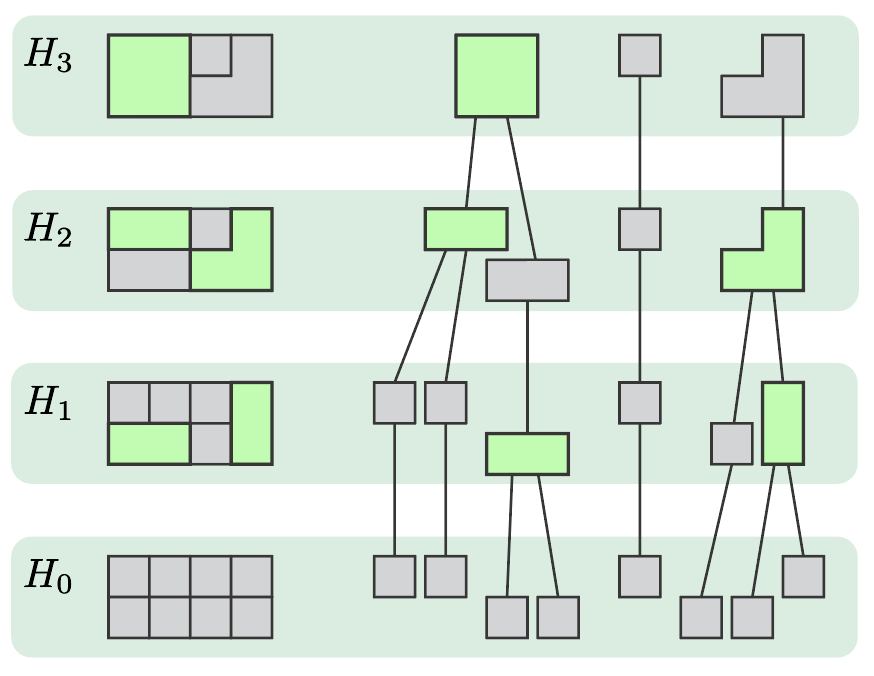}
        \caption{A four-level geographic hierarchy on a 4x2 grid (left), and the corresponding parent/child relationships (right). The coarse nodes that are merged at each level are highlighted in green.}
    \end{subfigure}
    \hfill 
    \begin{subfigure}[b]{0.48\textwidth}
        \includegraphics[width=\textwidth]{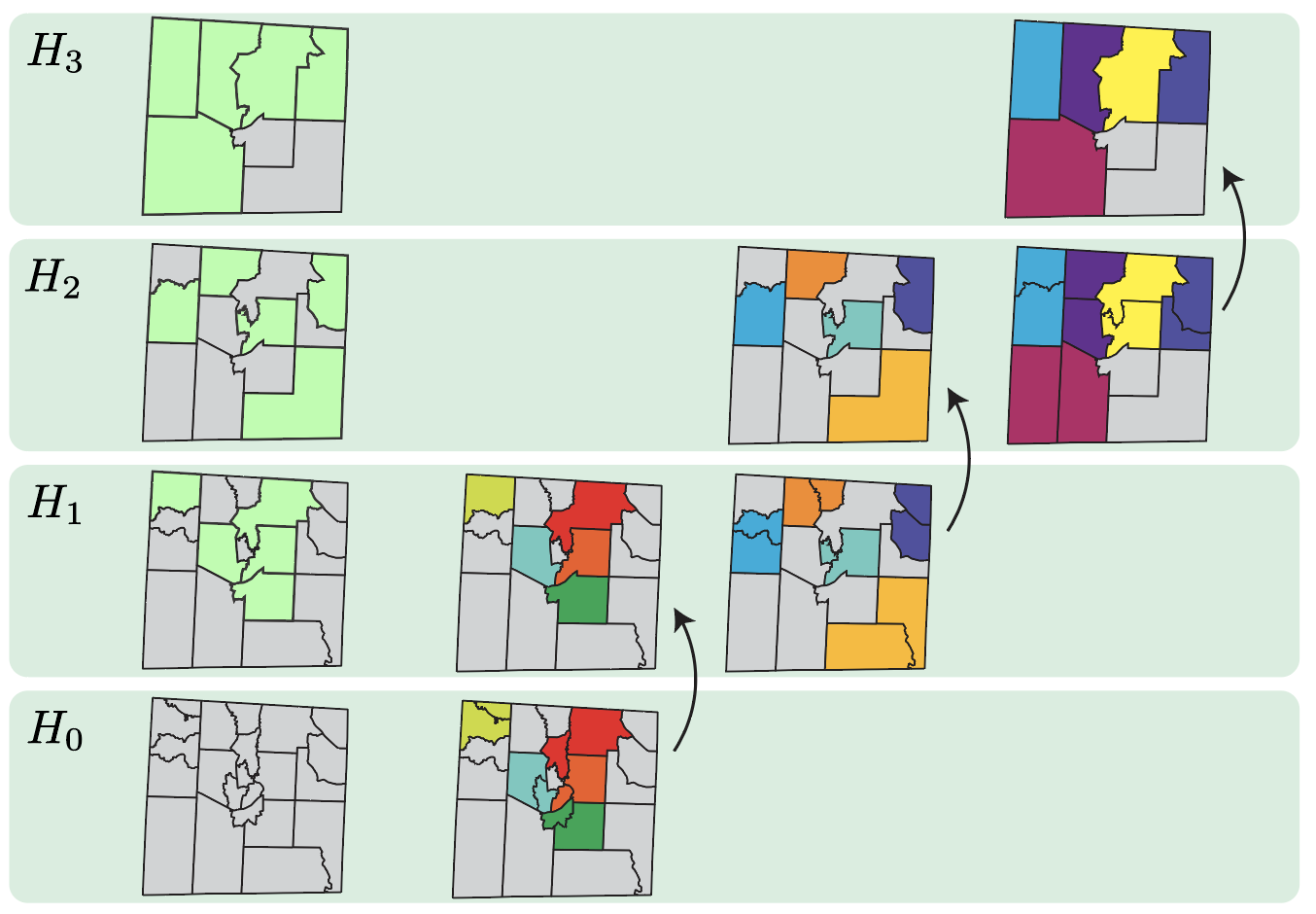}
        \caption{A four-level hierarchy on the precinct graph of Randolph County, North Carolina. Each level merges 5 pairs of nodes (indicated in color on the right), reducing a 22-node graph at $H_0$ to a 7-node graph at $H_3$. The coarse nodes that are merged at each level are indicated in green on the left.} 
    \end{subfigure}
    \caption{Geographic hierarchies on a 2x4 grid graph and the precinct graph of Randolph County, NC.}
\end{figure}

Our technique centers on a multiscale hierarchy drawn on top of the underlying graph $G$. We create a hierarchy $H = \langle H_0, H_1, H_2, \cdots H_L \rangle$, with $H_0 = G$ and each subsequent level $H_i$ formed by pairing and contracting edges on $H_{i-1}$ (or equivalently, merging pairs of nodes).  Not all nodes at $H_{i-1}$ must be a part of such a contraction, i.e., the matching need not be perfect nor maximal. This induces a forest-like structure, in which each node at level $i$ has one or two children at level $i-1$ and a parent at level $i+1$. 

When two nodes $(u,v) \in H_{i-1}$ are contracted into one node $w \in H_i$, we will say that $u$ and $v$ are \emph{merged} at level $i$, and that $w$ is \emph{split} at level $i-1$. As implied by the tree structure, we will say $u$ and $v$ are children of $w$, and $w$ is the parent of $u$ and $v$. 

An example geographic hierarchy and the corresponding forest structure is shown in Fig. 2(a), and a geographic hierarchy on the precinct graph of Randolph County, North Carolina is shown in Fig. 2(b). 

In practice, we consider several factors to make the specific hierarchy we choose particularly suitable for our method. We aim to have population-balanced nodes above a certain level of the hierarchy; we aim to limit the number of merged nodes per level; and we aim to have some notion of gradual change in how we choose which nodes to merge as we move up the hierarchy. We discuss these details in Section \ref{sec:hierarchy}.

\subsection{Parallel Tempering} 

%\begin{enumerate}
%    \item We use \PT (existing first paragraph)
%    \item What \PT is -- target measure, measure easy to sample from, interpolating measures, product measure, swaps via MH (show acceptance rate) (general) 
%    \item Fundamental problem in applying \PT to the hierarchy is that the spaces that $x_i$ and $x_i+1$ sample from are different, so we need some additional tool. (main idea is to highlight the problem)
%\end{enumerate}

The core of our technique is to use parallel tempering built on Metropolis-Hastings. We employ a single-node-flip proposal Markov chain independently on each level of the hierarchy, while proposing swaps between hierarchies at fixed intervals (i.e., exchanging the partitions at levels $H_i$ and $H_{i+1}$). Single-node-flip proposal chains and the Metropolis-Hastings algorithm are commonly used in the context of redistricting; for a brief discussion see Appendix~\ref{apdx:singlenodeflip} or, e.g., \cite{Chikina_Frieze_Pegden_2017,herschlag2020quantifying,fifield2020automated}. 

Parallel tempering  is a technique that uses several Markov Chains in parallel, each sampling from a different measure along with a swap mechanism to exchange states between different chains. One of these chains samples from a target measure and another samples from a measure space that is fast mixing. The remaining chains typically sample from measures that interpolate between the target measure (which may be very difficult to sample from) the fast-mixing measure (which may be far from the target). %The variable(s) being interpolated are commonly referred to as \emph{temperatures}, reflective of the technique's original use in [physics...] 
By sampling from all each of these measures in parallel, a Parallel Tempering sampler effectively samples from the \emph{product measure} on the joint state space across all chains; i.e., if the $i$th chain samples elements $x \in X$ from measure $\pi_i$, then the overall procedure samples the product $(x_1, x_2, \cdots x_L) \in X \times X \times \cdot \times X = X^L$ from the measure $\pi_1 \times \pi_2 \times \cdots \times \pi_L$. %\jcm{Why use $T$ here ? elsewhere number of levels is $L+1$. It is confusing to introduce $T$.}

\subsubsection{Extending the Typical Swap Procedure} 
Typically, Parallel Tempering also involves a \emph{Metropolis-Hastings swap} operation, which proposes a swap of states at two adjacent levels (i.e., moving from state $(x_1, \cdots x_i, x_{i+1}, \cdots x_T)$ to $(x_1, \cdots x_{i+1}, x_i, \cdots x_T)$). Critically in Parallel Tempering, the swap rate is designed preserve the marginal measure at each level. When all levels sample from the state space, the swap directly uses Metropolis-Hastings, which means that we accept the proposed swap with probability 
\begin{align}\label{eq:standardSwap}
A((x_i, x_{i+1}) \mapsto (x_{i+1}, x_i)) = \min\left(1, \frac{\pi_i(x_{i+1}) \pi_{i+1}(x_i)}{\pi_i(x_i) \pi_{i+1}(x_{i+1})}\right).
\end{align}

The swap proposal is particularly simple because  from the current state $(x_i, x_{i+1})$, one typically poses moving to a new state $(x_i', x_{i+1}')=(x_{i+1},x_i)$. However, we are free to pick $(x_i', x_{i+1}')$ in a probabilistic fashion given $(x_{i},x_{i+1})$. Specifically if $P\big((x_i, x_{i+1}) \mapsto (x_{i}', x_{i+1}')\big)$ is the probability of proposing $(x_i', x_{i+1}')$ from the curent state of $(x_i, x_{i+1})$ and the probability of accepting the proposal $(x_i', x_{i+1}')$  is
\begin{align}
A((x_i, x_{i+1}) \mapsto (x_{i}', x_{i+1}')) = \min\left(1, \frac{\pi_i(x_{i}') \pi_{i+1}(x_{i+1}')}{\pi_i(x_i) \pi_{i+1}(x_{i+1})}\frac{P\big(\,(x_i', x_{i+1}') \mapsto (x_{i}, x_{i+1})\,\big)}{P\big(\,(x_i, x_{i+1}) \mapsto (x_{i}', x_{i+1}')\,\big)}\right). 
\label{eq:PTprobmove}
\end{align}
then the product distribution $\pi_i \times \pi_{i+1}$ is preserved. One obtains \eqref{eq:standardSwap} from this formula by taking $P\big(\,(x_i, x_{i+1}) \mapsto (x_{i+}, x_{i})\,\big)=1$ and 0 otherwise. With this formulation, we are free to have $x_i$ and $x_i'$ live in one space and $x_{i+1}$ and $x_{i+1}'$ in another space. This is not possible with the standard parallel tempring swap.

\subsubsection{Overview of Our Swap Procedure} 
To employ  the classical  acceptance probability from \eqref{eq:standardSwap}, the state space must consist of products of the same space.  In our setting, the state space is $\Xi(H_0, d, J) \times \Xi(H_1, d, J_1) \times \cdots \times \Xi(H_L, d, J_L)$ where $J=J_0$ is the target measure specified in \eqref{eqn:scorefunc} and the subsequent $J_i$'s are a sequence of measures specified below in Section~\ref{ssec:highermeasures}. Recall that $\Xi(H_i, d, J_i)$ was defined in Section~\ref{sec:SamplingProblem}  to denote the space of valid $d$-partitions of graph $G$ subject to constraints imposed by a score function $J$. Observe that  $\Xi(H_i, d, J_i)$ and  $\Xi(H_{i+1}, d, J_{i+1})$ have different structure both because the graph $H_{i+1}$ is a coarsening of $H_{i}$ and because $J_i$ and $J_{i+1}$ might impose different constraints on the allowable redistricting plans. 
Nonetheless, one might expect $\Xi(H_i, d, J_i)$ and $\Xi(H_{i+1}, d, J_{i+1})$ are to be fairly similar, since  $H_{i+1}$ is obtained from $H_i$ by merging a handful of nodes and $J_i$ and $J_{i+1}$ are relatively close.

Hence, it is reasonable to hope that we can alter the redistricting plan $\xi_i$ on the graph $H_i$ into a redistricting plan $\xi_{i+1}'$ on $H_{i+1}$ and similarly $\xi_{i+1}$ into a districting $\xi_{i}'$ so that $(\xi_{i}', \xi_{i+1}') \in \Xi(H_i, d, J_i)\times \Xi(H_{i+1}, d, J_{i+1})$.

We will perform this deformation by introducing an auxiliary (time-inhomogeneous) Markov process  $\{(y_{\uparrow}^{(k)}, y_{\downarrow}^{(k)}) : k=0,1,\dots\}$ on $(\Xi(H_0, d) \cup \death) \times (\Xi(H_0, d) \cup \death)$ where $\death$ is a cemetery state described below. The process $(y_\uparrow, y_\downarrow)$ will be constructed so that if $(y_\uparrow^{(0)}, y_\downarrow^{(0)}) = (\xi_{i},\xi_{i+1}) \in\Xi(H_i, d, J_i) \times \Xi(H_{i+1}, d, J_{i+1})$ then after some number of steps $n$, determined by $(\xi_{i},\xi_{i+1})$, there is a positive chance that $y_\downarrow^{(n)} \in \Xi(H_{i}, d, J_{i})$ and $y_\uparrow^{(n)}\in \Xi(H_{i+1}, d, J_{i+1})$. Notice that $y\uparrow^{(n)}$, which started in the fine space, is now in the coarse space. The cemetery state $\death$ is added out of convenience as sometimes the Markov Chain $(y_\uparrow, y_\downarrow)$ will not have a reasonable move before the $n$th step. In such cases, we will set the state to $\death$,  signify that the process failed to produce usable results. The cemetery state  $\death$ is absorbing.  Setting $(\xi_{i}',\xi_{i+1}')=(y_\downarrow^{(n)}, y_\uparrow^{(n)})$,  when $(\xi_{i}',\xi_{i+1}') \in \Xi(H_{i}, d, J_{i}) \times \Xi(H_{i+1}, d, J_{i+1})$ we accept this proposal with probability
\begin{multline}
  A((\xi_i, \xi_{i+1}) \mapsto (\xi_{i}', \xi_{i+1}'))\\ = \min\left(1, \frac{\pi_i(\xi_{i}') \pi_{i+1}(\xi_{i+1}')}{\pi_i(\xi_i) \pi_{i+1}(\xi_{i+1})}\frac{P\big(\,(y_\uparrow^{(n)}, y_\downarrow^{(n)} ) \mapsto (y_\uparrow^{(n-1)}, y_\downarrow^{(n-1)} ) \mapsto \cdots \mapsto (y_\uparrow^{(0)}, y_\downarrow^{(0)} )\,\big)}
  {P\big(\,(y_\uparrow^{(0)}, y_\downarrow^{(0)} ) \mapsto (y_\uparrow^{(1)}, y_\downarrow^{(1)} ) \mapsto \cdots \mapsto (y_\uparrow^{(n)}, y_\downarrow^{(n)} )\,\big)} \right). \label{eq:swapsWPaths}
\end{multline}
and with probability 0 if $(\xi_{i}',\xi_{i+1}') \not\in \Xi(H_{i}, d, J_{i}) \times \Xi(H_{i+1}, d, J_{i+1})$. In particular if either of the $\xi_{i}'$ and $\xi_{i+1}'$ equals $\death$ then the probability of acceptance is zero.\footnote{We remark that equation \eqref{eq:swapsWPaths} differs from \eqref{eq:PTprobmove} in that we consider path probabilities rather than the probability of arriving at the given final state. One can see that this chain will preserve detailed balance by integrating (i.e. summing in this case) over all possible paths to a given proposal and noting that each of these paths respects detailed balance.}

\subsubsection{Swapping Schedule} 
To implement parallel tempering, we run the Markov chains at each level of the hierarchy for a specified number of $N$ steps. After the first $N$ steps, we then propose swaps between $H_0$ and $H_1$, $H_2$ and $H_3$, and so on; if $L$ is even we omit proposing an exchange with the state defined on $H_L$ at this step. After another $N$ steps we then propose swaps between $H_1$ and $H_2$, $H_3$ and $H_4$, and so on.  We omit proposing a swap with the state at $H_0$ at this step and also omit proposing a swap with the state $H_L$ if $L$ is odd. This scheme has been called the deterministic even-odd schedule and has been shown to have better mixing properties than stochastically determining swap directions \cite{syed2022non}.

\subsection{Swap Mechanism}
\label{ssec:SwapMech}
We now develop the details of the swap mechanism. In particular we construct the auxiliary chain $(y_\uparrow, y_\downarrow)$ that allows us to probabilistically swap between levels while targeting the appropriate state spaces.
% In this section, we present the coupled path construction to swap fine and coarse scales of the hierarchy. We also briefly discuss how to compute the forward and backward path probabilities and give more detail in Appendix~\ref{apdx:Swap}.
 The swap operation can be thought of a mutual random projection between the initial states and the final states $(y_\uparrow^{(0)}, y_\downarrow^{(0)}) \mapsto (y_\uparrow^{(n)}, y_\downarrow^{(n)})$. By ``mutual random projection'', we mean that after the states are swapped the previous fine state is projected to a coarse state and the previous coarse state is projected (or lifted) to a fine state. More exactly,  the fine partition, $\xi_i = y_\uparrow^{(0)}$, transitions from a district plan with tight population that splits coarsened nodes to a plan that does not split coarsened nodes and has looser population constraints, $\xi_{i+1}' = y_\uparrow^{(n)}$. This projection process is meant to keep the overall structure of $\xi_i$ and transfer it up to the coarse level, meaning that $\xi_i$ and $\xi_{i+1}'$ will be `close' to one another. The same is true for the relationship between $\xi_{i+1}$ and $\xi_i'$.

The random projection process will be a sequence of random moves that can be cast as a Markov Chain for which we can track path probabilities. At times, there will be no viable move in these chains at which point we send the chain to an absorbing state and reject the proposed swap.

The number of steps taken in the auxiliary chain is determined by the number of split coarse nodes of $H_{i+1}$ in the fine partition $\xi_i$ on $H_i$. Each step of the chain will reduce the number of split nodes in $y_\uparrow$ by one and increase the number of split nodes in $y_\downarrow$ by one. The fine to coarse chain gradually reduces the split nodes which allows $y_\uparrow^{(n)}$ to live at the coarsened level, $\Xi(H_{i+1}, d, J_{i+1})$. By splitting the same number of nodes in the coarse to fine projection $y_\downarrow$, the probability of running the process in reverse is non-zero and calculable, which is required in order to Metropolize to preserve the marginal measures.

In addition to gradually shifting the number of coarse nodes, we also gradually alter the constraints on the state space. We achieve this by interpolating the population constraints between the two energies $J_i$ and $J_{i+1}$.  Defining the fine and coarse population constraints as $\delta_i$ and $\delta_{i+1}$, respectively (with $\delta_i\leq\delta_{i+1}$), we interpolate between the spaces so that at the $k$th step in the chain the allowable population deviations for $y_\uparrow^{(k)}$ and $y_\downarrow^{(k)}$ are
\begin{align*}
\delta_\uparrow^{(k)} = \delta_i + k\frac{\delta_{i+1}-\delta_i}{n} \quad\text{ and } \quad \delta_\downarrow^{(k)} = \delta_{i+1} - k\frac{\delta_{i+1}-\delta_i}{n},
\end{align*}
respectively.

\begin{figure}[h]
    \centering
    \begin{subfigure}[b]{0.3\textwidth}
        \includegraphics[width=\textwidth]{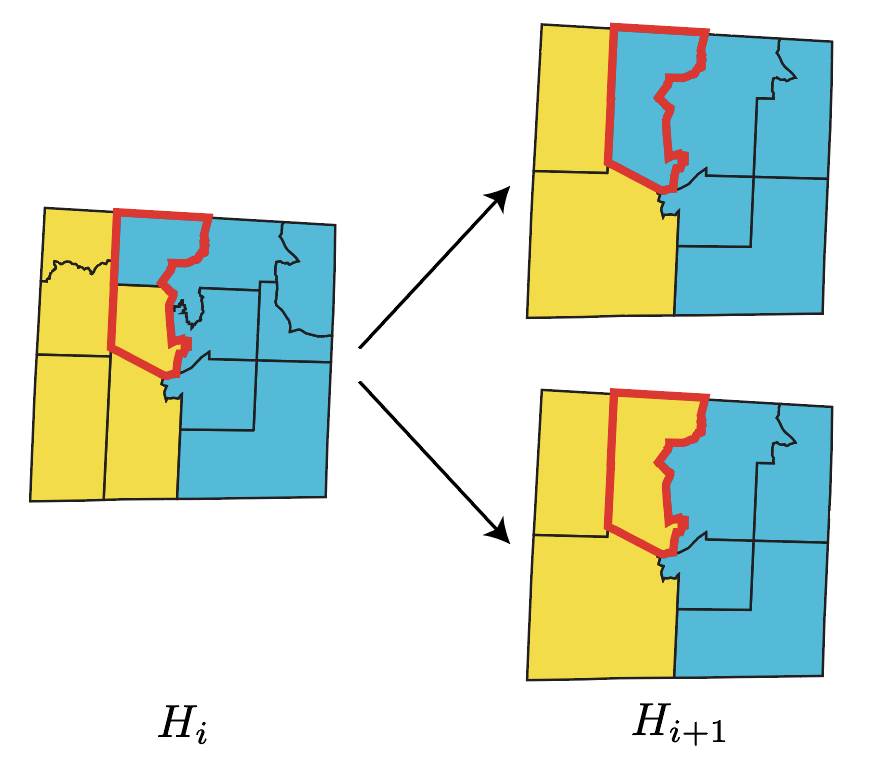}
        %\caption{The children of node $c_1$ and $c_2$, are assigned to two different districts in $\xi_1$, but $v$ must be assigned to only one district in $\xi_2$.}
        \caption{Node $v\in H_{i+1}$ is split in $\xi_{H_i}$, but must be assigned to only one district when being projected on to the coarser graph: either blue (top) or yellow (bottom).}
        %\caption{Coarse node $v$ covers two districts in $\xi_{F\to C}$, but must be assigned to only one district in $\xi_{F\to C}'$.}
        \label{sfig:mergenode}
    \end{subfigure}
    \hspace{0.1\textwidth} 
    \begin{subfigure}[b]{0.3\textwidth}
        \includegraphics[width=\textwidth]{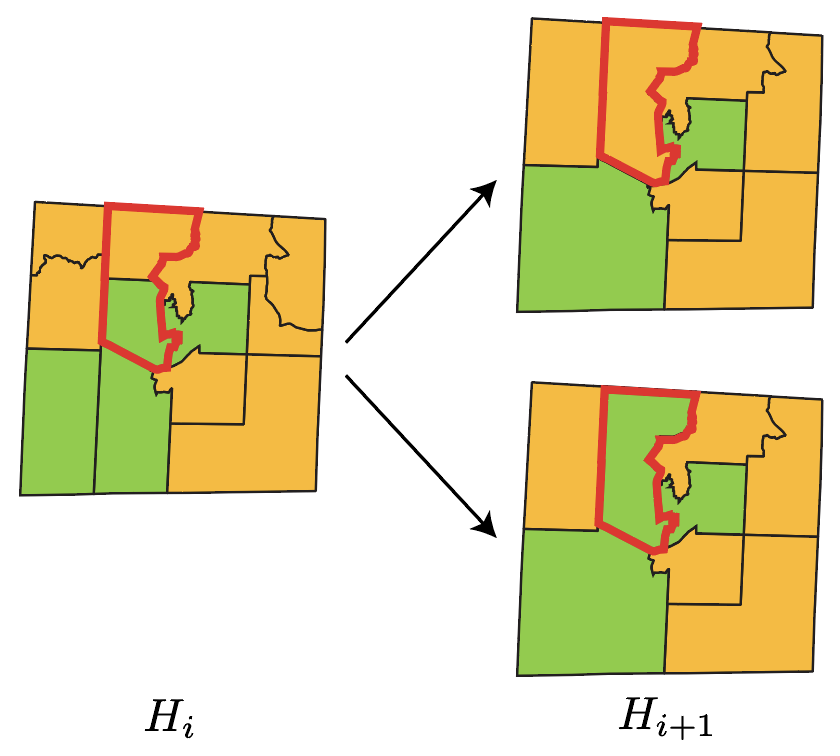}
        \caption{Node $v\in H_{i+1}$ is split in $\xi_{H_i}$, but un-splitting $v$ may be impossible. In this case it would disconnect either the green district (top) or the orange district (bottom).} 
        \label{sfig:cannotmergenode}
    \end{subfigure}
    %\hfill 
    %\begin{subfigure}[b]{0.3\textwidth}
    %    \includegraphics[width=\textwidth]{images/perverse_splitnode.png}
    %    \caption{} 
    %\end{subfigure}
    \caption{Part of the swap operation must project fine scale district plans, $\xi_F$, to some plan that can be represented on the coarse structure $H_C$. If a coarsened node in $H_{C}$ is mapped to different districts at the finer scale, we must resolve the fine scale assignment as we project the fine scale redistricting plan onto the coarsened space.}
\end{figure}

When merging nodes in $y_\uparrow$ we will uniformly choose a split coarse node to merge from the nodes which can successfully be merged.  For example, in Fig.~\ref{sfig:mergenode}, the highlighted coarse node $v$ has children $c_1$ and $c_2$, which are assigned to different districts (blue and yellow). When transitioning this configuration of $\xi_F$ to a coarse representation, $v$ must be assigned to either the yellow or blue district. 

In some cases, we will not be able to merge a split node in $y_\uparrow$ because merged node cannot be assigned to either color without disconnecting one of the districts; one such example is shown in Fig.~\ref{sfig:cannotmergenode}. In these cases, the Markov path within the swap operator will move to the death state $\death$. Similarly, if we cannot merge any node an remain under the new population constraints, $\delta_\uparrow^{(k)}$, we will also move to the death state $\death$. When splitting nodes in $y_\downarrow$, there may be no possible node to split that will lead to satisfying the population constraints given by $\delta_\downarrow^{(k)}$; in this case, we will also move to the death state $\death$.

%the fine-to-coarse partition un-splits a split coarse node (that is, it takes one of the $v$ described in 4.3.2 and flips the assignment of one of its children). Conversely, each change made to the coarse-to-fine partition splits a non-split coarse node (i.e., chooses a sibling pair $c_1, c_2$ that have the same assignment in $\xi_C$ and moves one of them to a different district). In this way, every move on the fine-to-coarse partition is exactly the reverse of a move on the coarse-to-fine one.

%i.e., it makes an change that is exactly the reverse of the change made to the fine-to-coarse partition.

\begin{figure}[h!]
    \begin{subfigure}[t]{\textwidth}
        \centering
        \includegraphics*[width=0.72\textwidth]{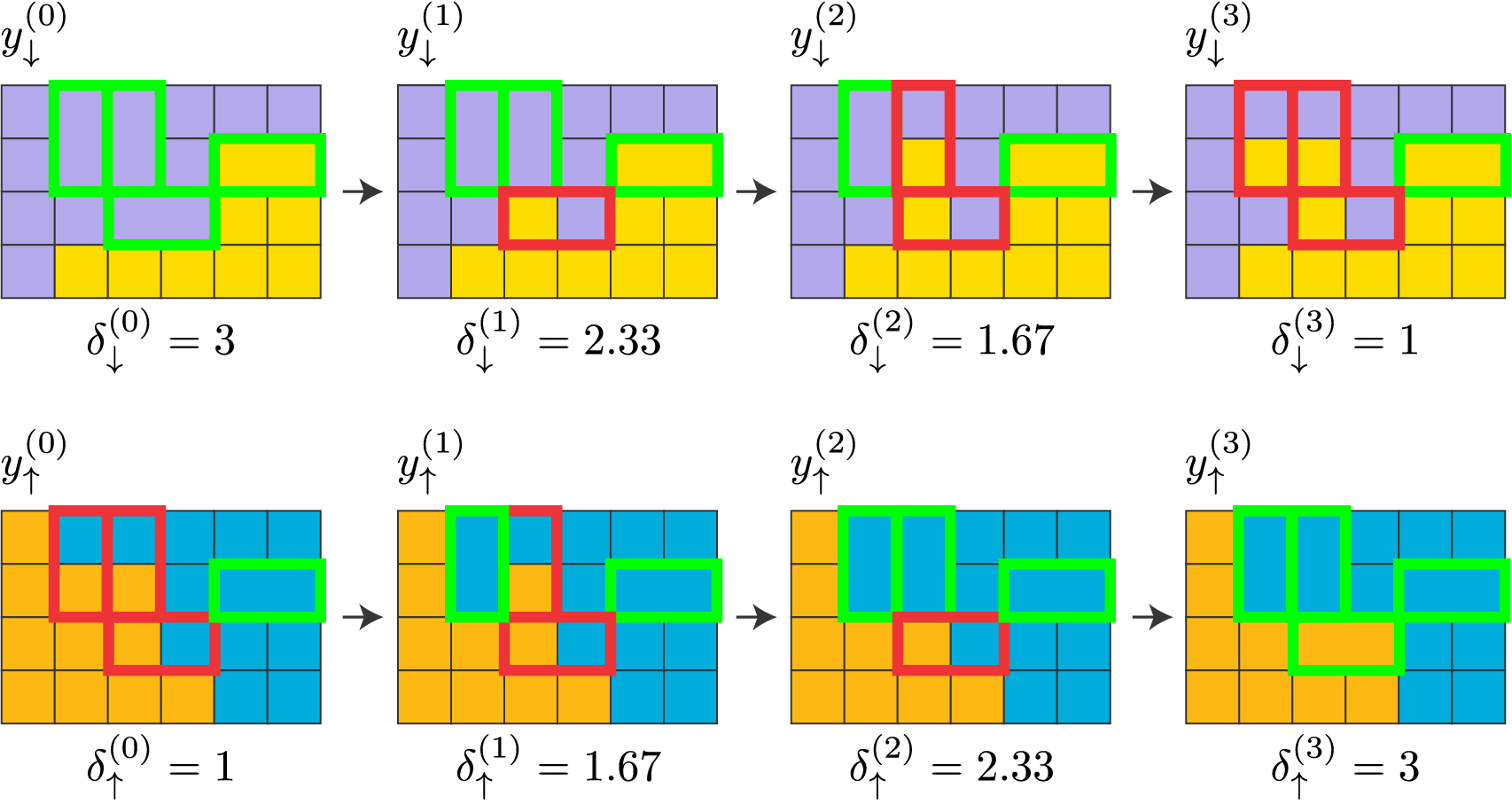}
        \caption{The swap mechanism applied on chain $(y_\uparrow, y_\downarrow)$. Split coarse nodes are marked in red outline; non-split coarse nodes are marked in green outline. There are three split coarse nodes in $y_\uparrow^{(0)}$, so three steps are taken. Each step on $\xi_\uparrow$ un-splits a coarse node (e.g., \orangesq\bluesq \, $\to$ \orangesq\!\orangesq \; for $y_\uparrow^{(0)} \to y_\uparrow^{(1)}$), and each step on $y_\downarrow$ splits a coarse node (e.g., \lavsq\!\lavsq $\to$ \yellowsq\lavsq\; for $y_\downarrow^{(0)} \to y_\downarrow^{(1)}$). Each intermediate step also keeps the district populations within the population bounds $\delta$ (linearly interpolated from $\delta_{i+1}=3$ at the coarse level to $\delta_i=1$ at the fine level). Notice that $y_\uparrow^{(3)}$ respects the coarse node boundaries of $H_{i+1}$, and $y_\downarrow^{(3)}$ respects the population bounds of level $i$.}
    \end{subfigure}
    \begin{subfigure}[t]{\textwidth}
        \centering
        \includegraphics*[width=0.9\textwidth]{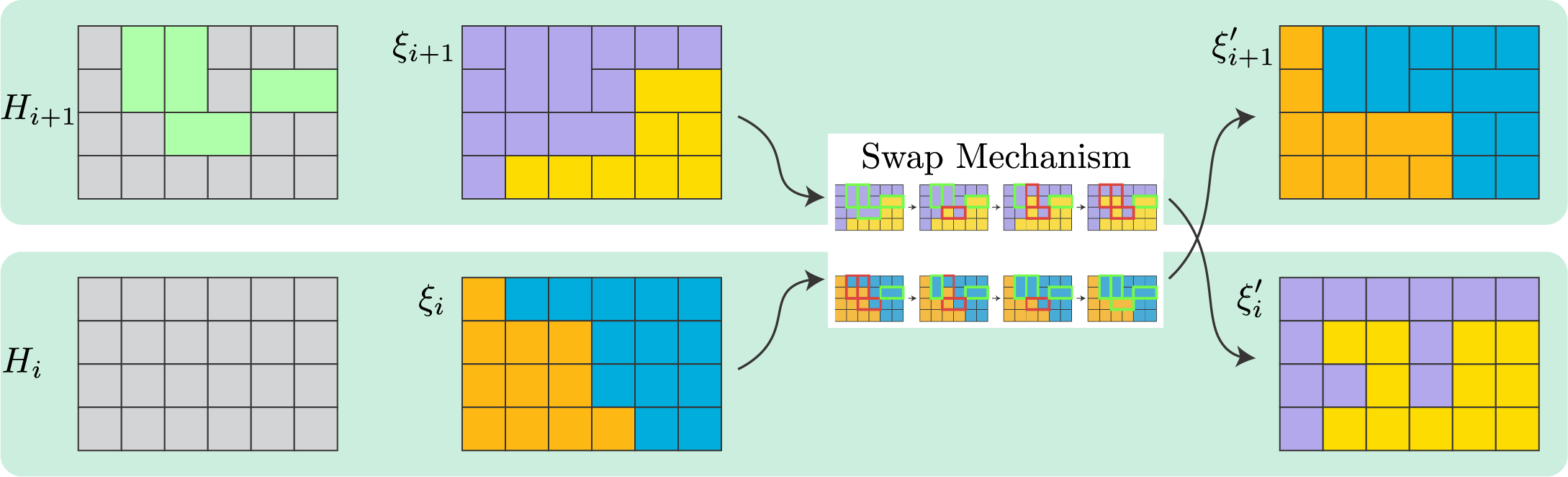}
        \caption{Left: coarse and fine hierarchy levels $H_{i+1}$ and $H_i$, with merged nodes highlighted in green. Right: the swap mechanism is used to transform districting $\xi_{i+1}$ into $\xi_i'$ and $\xi_i$ into $\xi_{i+1}'$, allowing us to swap them between levels in our parallel tempering scheme.}
    \end{subfigure}

    \caption{The swap mechanism transforms redistricting plans on adjacent levels in order to allow them to swap levels in the parallel tempering scheme.}
    \label{fig:swap_op}
\end{figure}

An example of swapping across scales on a 4x6 grid is shown in Figure \ref{fig:swap_op}. We also demonstrate how the swap fits into the large parallel tempering framework. 
% Below, we will refer to \emph{split coarse nodes}: these are nodes in $H_C$ whose two children are assigned to different districts (such as the red-outlined node in Figure 3(a)) by a given plan $\xi$. Similarly, \emph{non-split coarse nodes} are nodes in $H_C$ that have two children that are assigned to the same district by a given plan $\xi$. As described above, any valid partition of $H_C$ must have no split coarse nodes.

Designing the path probabilities
\begin{align*}
P\big((y_\uparrow^{(0)}, y_\downarrow^{(0)} ) \mapsto (y_\uparrow^{(1)}, y_\downarrow^{(1)} ) \mapsto \cdots \mapsto (y_\uparrow^{(n)}, y_\downarrow^{(n)} )\big),
\end{align*}
given in \eqref{eq:swapsWPaths} is, in principle, arbitrary; however, we have made several choices that appear to improve the acceptance ratio. When selecting a node to merge in $y_\uparrow$, we uniformly pick any split that has at least one valid single-node-flip move at the current state; once we've picked a split coarse node, if there are two possible moves, we then make a weighted choice to pick the one that leads to a more probable plan with probability two-thirds.%by placing a simple Bernoulli with parameter 2/3 in favor of the more probable state.
\footnote{We could also do this via tempering on the measure of interest. If we have two choices for $y_{\uparrow}^{(i), 1}$ and $y_{\uparrow}^{(i), 2}$, we would then weight our choice with this measure raised to some power.
% $\frac{\tilde{\pi}(\xi_{F\rightarrow C}^{i, 1})^\alpha}{\tilde{\pi}(\xi_{F\rightarrow C}^{i, 1})^\alpha + \tilde{\pi}(\xi_{F\rightarrow C}^{i, 2})^\alpha}
% $
% where $\alpha$ is an arbitrary tempering parameter typically chosen within the interval $[0,1]$. For the current work, we have just made the simple choice above.  Here $\tilde{\pi}$ would be some interpolated measure that accounts for differences in the weights and constraints.
}

When selecting a node to split in $y_\downarrow$ move we weight all possible valid splits by the their probability, i.e. by $\pi(\xi_{C\to F; i}^j)^\alpha$ where $\alpha$ is a tempering parameter and $i$ is indexed over the possible moves when merging the $(j+1)$th node in $\xi_{C\to F}^j$. The parameter $\alpha$ is taken to be 0.1 in our results below, but it appears that there is significant flexibility in this choice.

In the above, there are many other possible algorithmic choices here and we make no claims of optimality. Although we do not report them exhaustively, we have found good mixing results in a variety of informed algorithmic choices. Heuristically, we find success when we control the probabilities so that (i) projections remain reasonably compact and are consistent with the samples found at the complementary level and (ii) the forward and backward chain probabilities tend to be reasonably close. 

% Further details can be found in Appendix \ref{apdx:Swap}. 

% An example of a three-step \textsc{PairedExchange} swap is given in Figure \ref{fig:swap_op}.

% \section{Practical choices/Implementation/idk}
% this section needs a better title 
% Idk, we had the below 2 subsections as parts of the previous section but they feel kind of different. the previous section is about the sampling mechanism, whereas these two are a bit more specific to our instantiation of it. 

% Our sampling method is agnostic to the choice of measure and the choice of hierarchy. In this section, we discuss the specific policy-driven choices we make with respect to the tempered measures and the construction of the hierarchy. 

\subsection{Construction of the Higher-Level Measures}
\label{ssec:highermeasures}
In any parallel tempering scheme, one needs to construct a family of measures that interpolates between a target measure and one that mixes (relatively) quickly. In this section, we describe how we construct a sequence of measures for each level of the hierarchy for our example measure of interest. 

In the current work, we consider a measure with population constraints and a preference for compact plans as defined by the Polsby-Popper score. As we coarsen the hierarchy, the nodes become more populated and thus we need to relax the population constraints. Similarly, we relax the compactness weight so that districts can mix more readily. We index the sequence of measures by $J_\ell$ for $\ell\in\{0,...,L\}$. The zeroth measure will be the target measure and simply be referred to as $J$ (i.e. $J=J_0$) and take on the same form as \eqref{eqn:scorefunc}.  The subsequent measures will also take on the same overall form as \eqref{eqn:scorefunc}, however they will use different parameters in the population and compactness scores so
\begin{align}
J_\ell(\xi) = J_{\text{contiguous}}(\xi) + J_{\text{pop}}(\xi; \delta_\text{pop}^{(\ell)}) + w_{comp}^{(\ell)} J_\text{comp}(\xi).
\end{align}

For the population constraint, we scale the allowable population bound at level $\ell$ with the average population of the nodes at that level
\begin{align}
\delta_\text{pop}^{(\ell)} = \alpha \times \frac{\text{total state population}}{\text{number of districts} \times |V(H_\ell)|},
\end{align}
where $\alpha$ is fixed such that the population bound at the finest level is the desired population variation in the ensemble (for example, 2\%). In practice, this means that the allowable population deviation at each level is approximately a constant number of nodes, allowing significant border exchanges using single node flips (regardless of the size of the nodes).  The allowable populations at level $\ell$ are
\begin{align}
\frac{\text{total state population}}{\text{number of districts}} \pm \delta_\text{pop}^{(\ell)},
\end{align}
as seen in \eqref{eq:popconstraint}.

Second, we relax the compactness weight, $w_{comp}^{(\ell)}$
% the measure at level $i$ of the tempering scheme is 
% \[P_i(\xi) \propto \exp\left(-\kappa_i \sum_{d \in \xi} \frac{\text{perim}(d)^2}{\text{area}(d)}\right) \]
by decreasing it (i.e., reducing the impact of compactness) at higher levels of the tempering scheme. This allows less-compact plans at coarser levels. Empirically, we find that a hyperbola-shaped curve for $w_{comp}^{(\ell)}$ (as a function of $\ell$; flatter at the finer levels, close to linear at coarser levels) results in a high amount of swapping between levels. Specifically, for a specified target weight $w_{comp} = w_{comp}^{(0)}$ and $\ell \in \{1,...,L\}$ for each other level in the hierarchy, we take
\begin{align}
w_{comp}^{(\ell)} = 1 + w_{comp} - \sqrt{1 + (2 w_{comp} + w_{comp}^2)\left(\frac{\ell}{L}\right)^2}.
\end{align}
The above formula ensures that the weight at the coarsest level is 0 (i.e., the measure at the coarsest level of the scheme is uniform on loosely population-balanced partitions).  In general, uniform partition spaces without any compactness considerations form percolating graphs with boundaries that are close to space filling curves \cite{najt2021empirical}.  However, the graph at the coarsest levels is designed to have only a few nodes per district and thus we do not encounter this issue. This choice of tempering scheme is empirically driven; other schemes are also likely to work well.

% [Maybe a little figure showing the hyperbola curve we use for Iowa/CT and/or the pop bounds curve]. 

\subsection{Generation of the Hierarchy}
\label{sec:hierarchy}
Our sampling procedure could, in principle, work on a wide class of hierarchies. However, we have found that the choice of hierarchy may affects the mixing time of the parallel tempering scheme. In particular, the hierarchy may impact the Metropolis-Hastings acceptance ratio of the swap operation. %, and one could concievably attempt to construct such a hierarchy by hand. However, given the scale of our graphs, 
In practice, we generate the hierarchies computationally, based on the guiding principles outlined below.

We prioritize graph contractions that prioritize target node populations and seek to mitigate abrupt changes to the compactness score. We also ensure that contractions will not prevent mixing and explain how we achieve this in this section. We then iteratively add levels to the hierarchy by repeatedly sampling $k$ disjoint edges on the current coarsest level. We contract these edges to form a new coarser graph that becomes the new top of the hierarchy. The edges are chosen according to an edge weight function $w_\ell(u,v)$, which encodes how desirable it is to merge nodes $u$ and $v$ at level $\ell$:
\begin{align} 
w_\ell(u,v) = w_\ell^{(pop)}(u,v) +  w_\ell^{(comp)}(u,v) + w^{(articulation)}(u,v),
\end{align} 
where the three terms on the right hand side correspond to the population weight, compactness weight, and whether or not the merge will generate an articulation point. We expand and define each of these elements below.

We then sort the edge weights from largest to smallest and greedily select $k$ edges by selecting the edge with the largest weight (i.e. most desirable) such that the edge does not share a node with an edge that has already been selected. We note that this algorithm may not always be able to find $k$ edges; if this is the case, then we just select however many edges we are able to with this procedure. 

This process is deterministic. It is also possible to form different hierarchies by selecting a random number between zero and $w_\ell(u,v)$ for each edge and sorting these numbers instead of the weights. In general, we believe there are many feasible algorithms to contract edges so long as the guiding principles discussed below are followed. We give details on the properties of the hierarchies in Appendix~\ref{apdx:hierarchydetails}.

\subsubsection{Number of Merged Nodes Per Level}
Each step of the projections involves a probabalistic event of selecting the next nodes and then merging/splitting them. The forward and backward probabilities are computed by taking a product of the probability of these events. Thus we would like to reduce deviation of the ratio of these probabilities away from one since large deviations can have a significant impact on the acceptance ratio found in \eqref{eq:swapsWPaths}. The number of merges and splits will depend on (i) the number of merged nodes between levels $H_i$ and $H_{i+1}$, and (ii) the relative size of the boundary nodes to interior nodes (which will depend on the size of the graph and the number of partitions). In this work, we choose to keep the number of merged nodes at each level (roughly) constant with $k=30$. This results in the number of levels in the hierarchy to scale linearly with the number of nodes in the base graph. In practice, state-level precinct graphs range in size from several hundred to several thousand, so this linear dependence is quite reasonable and leads to a few dozen levels in the hierarchy in many cases. Again, fixing this constant may not be optimal and there may be reason to let the target number of merged nodes vary with the level; we do not seek to improve upon the current scheme here as we have found the above parameter capable of efficiently sampling our examples of interest.

\subsubsection{Population} 
One potential barrier to the mixing of single-node-flip MCMC samplers is the relatively large range of population in precincts in a given state. For example, Connecticut's precincts range in population from 53 to 26,970; this largest precinct is 3.7\% of a congressional district's ideal population, which is quite high relative to an desired population variation of, for example, 2\%. The district assigned to such a large-population node would almost never be flipped by a single-node-flip chain alone: such a flip would almost always take one or both affected districts out of the population bounds\footnote{In this example, one could concievably flip the 3.7\%-sized node from $d_1$ to $d_2$ if $d_1$ was 1.9\% over ideal population and $d_2$ was 1.9\% under ideal population; anything closer to ideal would result in one or both districts going outside the bounds. If the desired population bounds were 1.5\% instead, the large node would never flip.}.

To account for this, we aim to gradually regularize our node populations at higher levels of the hierarchy. Once we have constructed $H_i$ and are constructing $H_{i+1}$, we define the target node population, $p_\ell$, in the hierarchy as the total graph population 
\begin{align}
p_\ell = \frac{\text{total population}}{|V(H_i)|-k}
\end{align}
and then define the population element of the edge weights to be 
\begin{align}
w_\ell^{(pop)}(u,v) = -\alpha^{(pop)}_\ell \times \max\left(\frac{\text{pop}(u,v)}{p_\ell}, \frac{p_\ell}{\text{pop}(u,v)}\right)
\end{align}
for a tunable positive constant $\alpha^{(pop)}_\ell$ that we can vary by level. 

\subsubsection{Flat Compactness} 
We have found that the swap operation works best when the compactness of the nodes at each level remains fairly constant. This may, at first, be somewhat counter-intuitive: One may imagine that having very compact coarse nodes would be preferable. The reason for keeping the compactness scores similar across levels is because significant spikes or dips in compactness during the projection operations make it less likely that the swap mechanism will mutually project the districts into regions with compactness scores that are comparable to the current state of the chain. 

For example, consider two adjacent nodes that when joined are rectangular, but that are split by a fractal-like river (i.e. a jagged boundary with high perimeter). A swap operation attempting to split those nodes (i.e. from a coarse to fine representation) would lead to a drastic increase in the perimeter for the two districts that would share the boundary, and hence a drastic decrease in compactness. If the fine scale typically keeps the two nodes in the same district simply due to the energy of the distribution, then forcing this split would reduce the probability of accepting the proposed swap.

To ameliorate these sudden changes, we seek to merge nodes such that the merged node has a similar compactness score as its two child nodes. Formally, if $u\in V(H_{i+1})$ is a node formed by merging $c_1$ and $c_2\in V(H_i)$, then we scale the weight via
\begin{align}
w_\ell^{(comp)}(c_1,c_2) = -\alpha_\ell^{(comp)} \times \max\left(\frac{J_{comp}(u)}{\frac12 (J_{comp}(c_1) + J_{comp}(c_2))}, \frac{\frac12 (J_{comp}(c_1) + J_{comp}(c_2))}{J_{comp}(u)}\right)
\end{align}
for a tunable positive constant $\alpha^{(comp)}_\ell$ that we can vary by level. 

\subsubsection{Articulation Nodes} 
We disallow merging any edges where merging the nodes would lead to an articulation node in the graph. Recall that an articulation point is a node that, if removed, would disconnect the graph. Articulation points create hard walls in the state space with respect to single node flip moves: An articulation point makes it impossible to moved certain nodes from one district to another without making one district discontiguous. We therefore set the articulation weight to zero if the merge does not create an articulation point and $-\infty$ otherwise. See Appendix~\ref{apdx:singlenodeflip} for more discussion of articulation points and how they prevent mixing when flipping a single node at a time. 

\subsubsection{Other Considerations} 
In the current work, we have focused only on population, compactness, and contiguity. In general, one may wish to make other considerations such as county/community/municipal preservation, or to sample from plans with majority-minority districts. Although we do not extend our methods to these considerations in these work, we believe that given the small local changes made at each level along with the ability to relax conditions at coarse hierarchical levels, that our algorithm may nicely extend to these additional measures. We defer such investigation for future work.

\section{Numerical Results}
\begin{figure}
\begin{center}
\includegraphics[width=0.45\textwidth, clip=true, trim=4cm 5.5cm 3cm 5cm]{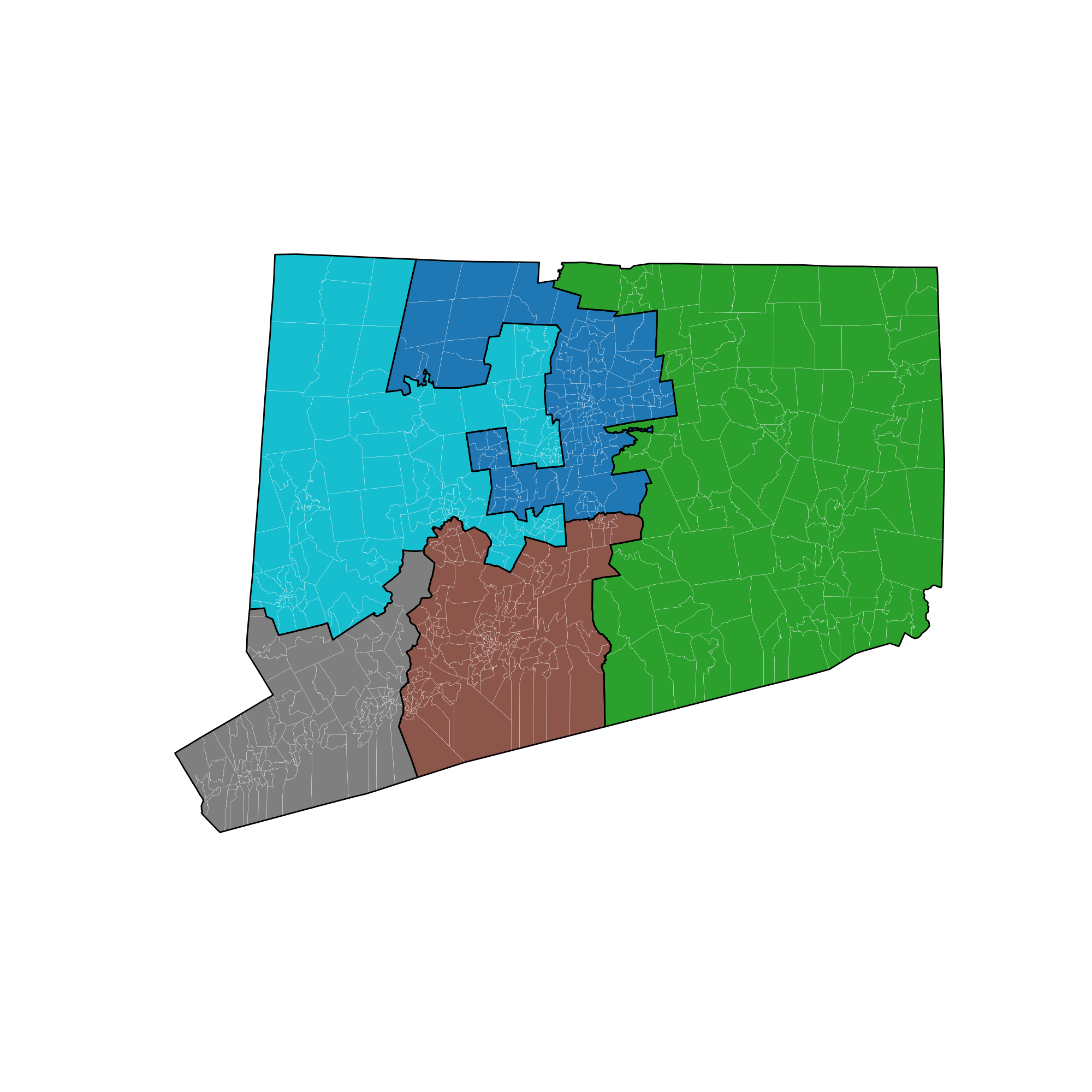}
\end{center}
\caption{We show the congressional districts and precincts from the 2021 Connecticut redistricting cycle. Data comes from the VEST project \cite{CTvest} via the Redistricting Data Hub \cite{redistrictingDataHub}.}
\label{fig:CTpct}
\end{figure}

We implement and test our sampler on the precinct graph of Connecticut (CT) displayed in Figure~\ref{fig:CTpct}. Connecticut has 739 precincts that are split into 5 congressional districts. Some of the precincts are multi-polygonal which means that they can potentially lead to the creation of discontiguous districts. To remedy this, we merge them with other precincts so that the joined unit forms a contiguous unit (see Appendices~\ref{apdx:singlenodeflip}~and~\ref{apdx:hierarchydetails} for details). After this joining process we have 695 precincts (or merged precincts) which serve as the nodes in our graph. Adjacency in this case is defined via rook adjacency: nodes are adjacent if they share non-zero length borders. This leads to the 695 nodes of the graph being connected by 1942 edges.

We remark that this is significantly larger than any single-node-flip algorithm has been shown to converge on in the past. To our knowledge the previous largest graphs that have been shown to successfully mix using these dynamics are on Wake County of North Carolina with 202 nodes and on Mecklenburg county in North Carolina with 194 nodes; the successful samplers were done with 11 state house districts and 5 state senate districts in Wake\footnote{The state senate districts actually were done on 203 nodes in which all of Franklin county was added as a single extra node since it was mandated to be unsplit.} and 12 state house districts and 5 state senate districts in Mecklenburg \cite{jcmReport,jcmRebuttal}. We also remark that we have confirmed we do not see mixing in Connecticut when running as single chain at the target measure.

% are multi-polygonal wand 1942 edges, and has five congressional districts. %Iowa has 1680 precincts and 4021 edges, and has four congressional districts. 
% Some of these precincts are multi-polygonal; since districts must be contiguous the precincts that connect these multi-polygonal units must be in the same district as the mult
% [Something about these graphs being much larger than demonstrated in the past?]. 

\subsection{Run Parameters, Initialization, and Hardware} 
We sample the 5 congressional districts in Connecticut with a target distribution that contains no more than a 2\% population deviation for each district. We also tune $w_{comp}$ so that it is (roughly) comparable to the enacted distribution and use a weight of $w_{comp}=0.8$ . We also vary this weight below in several different ways to investigate the robustness of our results to different choices of compactness and to demonstrate the ability to tune and focus the compactness levels on different regimes. 

To test for convergence, we examine a set of four independent runs. Each run uses the same 24-level hierarchy that merges approximately 30 nodes at each level. We launch a fifth run with a different, 22-level hierarchy to verify that the choice of hierarchy does not influence our samples at the target distribution.  Details of the hierarchies can be found in Appendix~\ref{apdx:hierarchydetails}.

On each run, each chain begins at a unique initial condition that is a valid redistricting of the state space $\Xi(H_i, d, J_i)$. The initial condition is chosen by first drawing a random spanning tree on the graph and cutting it so that there is one district within the population bounds and that the remaining nodes in the graph can be feasibly divided into 4 districts with in the population constraints. This district is fixed and we then repeat this process recursively on the remaining nodes. If we cannot find a district based on the tree, we make up to 100 attempts by drawing new trees.  If all attempts fails, we start the process again on the whole graph. This process is similar to ignoring the filter weights in the sequential Monte Carlo sampler presented here \cite{mccartan2020sequential}, and is identical to the initialization procedure presented used in Metroplized forest Recombination sampler \cite{autry2023metropolized}.  

We launch each run on an AMD Ryzen 9 5950X 16-Core Processor that is hyper-threaded to effectively contain 32 cores. We run the chains for 12 million single-node-flip steps, and propose parallel tempering swaps every 30 steps. The runs took approximately 11 hours on this hardware.  

%\subsubsection{Iowa} 
%We created a 32-level hierarchy according to the heuristics described in Section \ref{sec:hierarchy-gen} and Appendix ? on the 1680-node precinct graph of Iowa, with approximately 50 nodes merged per level and 16 nodes at the coarest level. At the finest level of the parallel tempering scheme, we set a population deviation bound of 2.0\% and a compactness energy using the isoperimetric (Polsby-Popper) score with a weight of 0.7. 
%We ran each chain for 12 million single-node-flip steps, with parallel tempering swaps proposed every 30 steps, which took approximately 40 hours on the hardware described above. 

% \subsection{Low-Dimensional Observables} 

\begin{figure}[h!]   
    \begin{subfigure}{0.45\textwidth}
        \includegraphics*[width=\textwidth]{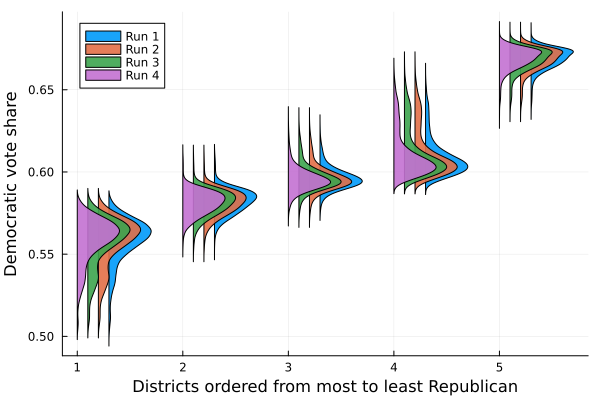}
        \caption{The ordered marginal distributions for the percent of the vote
        received by the Democrat in each of the four Connecticut runs. After 12 million steps, the distributions are extremely similar. The maximum pairwise total variation averaged across the marginal distributions is 0.048.}
        \label{sfig:voteCTfinal}
    \end{subfigure}
    \hfill 
    \begin{subfigure}{0.45\textwidth}
        \includegraphics*[width=\textwidth]{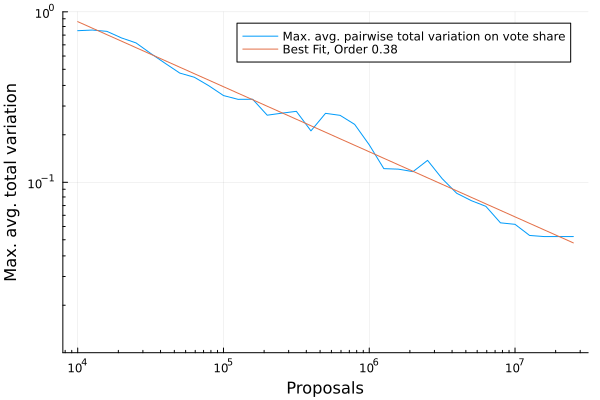}
        \caption{Convergence of the marginal vote share of the four Connecticut runs. The maximum average pairwise total variation decreases according to a power law with order 0.38.} 
        \label{sfig:voteCTconvergence}
    \end{subfigure}

    \begin{subfigure}{0.45\textwidth}
        \includegraphics*[width=\textwidth]{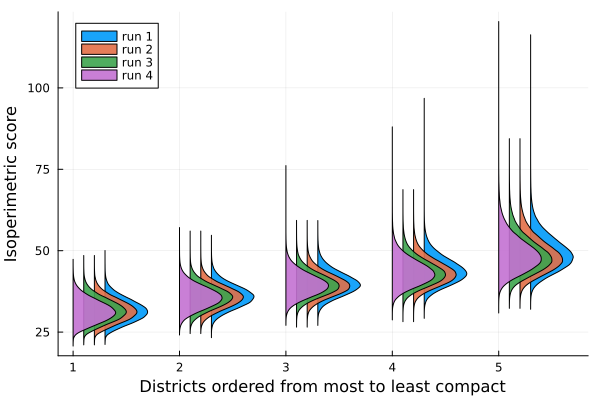}
        \caption{The ordered isoperimetric ratios of each district in each of the four Connecticut runs. After 12 million steps, the distributions are extremely similar. The maximum pairwise total variation averaged across the marginal distributions is 0.035.} 
        \label{sfig:isoCTfinal}
    \end{subfigure}
    \hfill 
    \begin{subfigure}{0.45\textwidth}
        \includegraphics*[width=\textwidth]{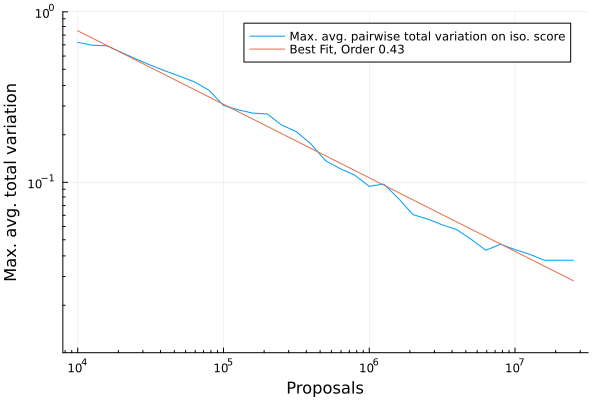}
        \caption{Convergence of the isoperimetric ratios of the four Connecticut runs. The maximum average pairwise total variation decreases according to a power law with order 0.43.} 
        \label{sfig:isoCTconvergence}
    \end{subfigure}
    \caption{Convergence on observables of interest of four Connecticut runs. For all pairs of runs, we average the total variation across all of the marginals displayed in Figures~\ref{sfig:voteCTfinal}~and~\ref{sfig:isoCTfinal} and then examine the maximum average distance across all pairs.}
    \label{fig:convergence}
\end{figure}

\subsection{Observables}
We analyze two low-dimensional observables of interest.
The first observable that we use to evaluate the collected ensemble is the \emph{relative vote share} of each district. That is, for each district in a generated map, we compute the relative number of Democratic and Republican votes that were cast in the precincts contained within that district in a given election, defined by
\begin{align}
\frac{V_{Dem}(\xi^{(i)})}{V_{Dem}(\xi^{(i)}) + V_{Rep}(\xi^{(i)})},
\end{align}
where $V_{Dem}(\xi^{(i)})$ are the number of votes for the Democratic candidate in district $\xi^{(i)}$ and $V_{Rep}$ is the same for the Republican votes.
Any election or set of elections can be used for this evaluation; in this work, we use the results from the 2020 Presidential election. 

Specifically, we order the vote shares of each district within a plan, from the least Democratic to the most Democratic; for each rank, we consider the distribution of the vote share margin. This allows the ensemble to not only capture the range of variation of number of seats won, but also the variation in the margin of victory of each seat. 

The secondary observable that we evaluate is the isoperimetric ratio of each district which is the scaled inverse of the Polsby-Popper score (i.e. it is mathematically equivalent to look at either the Polsby-Popper score or the isoperimetric ratio). We order the scores of each district from most to least compact, and consider the distribution of compactness scores at each ordered rank.

We select these two sets of observables as the first demonstrates what the expected range of compactness whereas the second can be used to understand the typical partisan effect when changing the policy expressed by the distribution.

\subsection{Convergence} 
The distributions of vote share margins and isoperimetric ratios in our ensembles for Connecticut (at compactness weight $w_{comp}=0.8$) are shown in Figs.~\ref{sfig:isoCTfinal}~and~\ref{sfig:voteCTfinal}. Visually, we find very close agreement between the marginal distributions and formalize this statement below. 

% To measure the error in the vote share margins we build histograms on each vote share margin with a bin width of 2\% and a bin division centered at 50\% (e.g. one bin is 50-52\%). For each of the four runs, we compare it everyone of the other runs by averaging the total variation across the 5 rank ordered histograms. We then take the maximum error across all run comparisons and display this error as a function of the number of steps in Fig.~\ref{fig:convergence} (b,d).    %For the purpose of this work, we note that this particular observable was chosen simply to demonstrate our algorithm; one can just as easily compute measures of compactness, responsiveness (e.g., efficiency gap), etc. on the generated ensemble. We omit such studies in the present work, as we are primarily interested in demonstrating the effectiveness and convergence of our algorithm. 

% We use the vote share margin and isoperimetric scores to evaluate the convergence of our chains. In particular, 
We compute the total variation distance between each pair of the four runs at the target distribution.  For the vote shares we use histograms with a bin width of 0.2\% with a bin division centered at 50\% (i.e. one of the bins is 50\%-50.2\%, another at 50.2\%-50.4\%, ...) and a bin width of 0.5 for the histogram on the isoperimetric ratios with a bin division at 0 (i.e. one of the bins is 0-0.5, another from 0.5-1, ...). For any two runs, we examine the total variation across all 5 rank-ordered marginal distributions and then take the average of these distances. We take the maximum averaged total variation across all pairs of chains and plot the maximum averaged total variation as a function of the number of proposals in Figures~\ref{sfig:voteCTconvergence}~and~\ref{sfig:isoCTconvergence}. 

Using the vote share margin, the average total variation decreases (roughly) according to a power law with order 0.32. We achieve an averaged total variation of 4.8\%. Using the isoperimetric score, the average total variation decreases according to a power law with order 0.43; we achieve an averaged total variation of 1.7\%. Our results strongly suggests that the chains have converged to the sampled measure on the low dimensional observables of interest.

% {\color{red} How hard would it be to report expected round trip times for each of the runs?} 

\subsection{Hierarchy Independence}

We verify that the samples produced by our sampler do not depend on the particular choice of hierarchy by comparing the observables of runs that used different hierarchies. In particular, we targeted the measure described above (with population deviation of 2\%, compactness weight of 0.8) using a different, 22-level randomized hierarchy, and compare the marginal vote share and the isoperimetric scores to confirm that both runs converge to the same distribution.

\begin{figure}[h!]   
    \begin{subfigure}{0.45\textwidth}
        \includegraphics*[width=\textwidth]{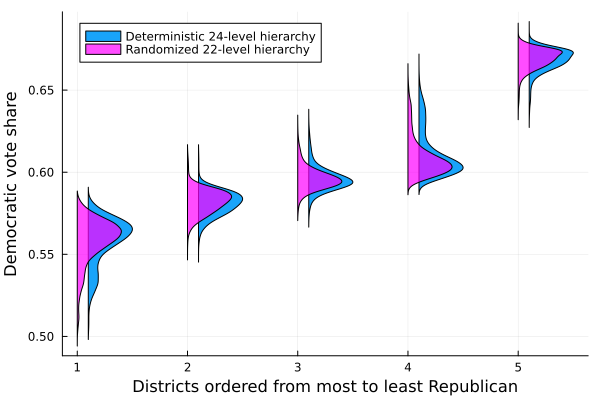}
        \caption{The ordered marginal distributions for the percent of the vote
        received by the Democrat in runs using two different hierarchies. After 12 million steps, the maximum pairwise total variation averaged across the marginal distributions is 0.054.}
        \label{sfig:voteHierarchyIndep}
    \end{subfigure}
    \hfill 
    \begin{subfigure}{0.45\textwidth}
        \includegraphics*[width=\textwidth]{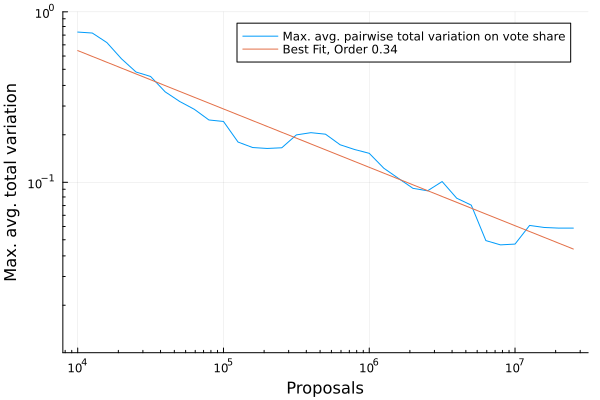}
        \caption{Convergence of the marginal vote share in runs using two different hierarchies. The maximum average pairwise total variation decreases according to a power law with order 0.34.} 
    \end{subfigure}

    \begin{subfigure}{0.45\textwidth}
        \includegraphics*[width=\textwidth]{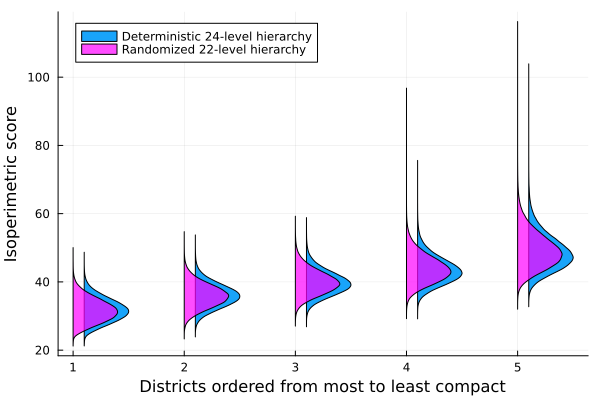}
        \caption{The ordered isoperimetric ratios of each district in runs using two different hierarchies. After 12 million steps, the maximum pairwise total variation averaged across the marginal distributions is 0.035.} 
        \label{sfig:isoHierarchyIndep}
    \end{subfigure}
    \hfill 
    \begin{subfigure}{0.45\textwidth}
        \includegraphics*[width=\textwidth]{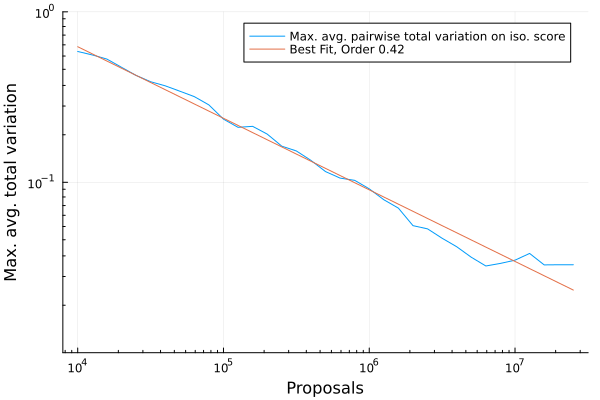}
        \caption{Convergence of the isoperimetric ratios in runs using two different hierarchies. The maximum average pairwise total variation decreases according to a power law with order 0.42.} 
    \end{subfigure}
    \caption{Convergence on observables of interest on runs using two different tempering hierarchies: a deterministically-constructed 24-level hierarchy and a randomly-constructed 22-level hierarchy. We average the total variation across all of the marginals displayed in Figures~\ref{sfig:voteHierarchyIndep}~and~\ref{sfig:isoHierarchyIndep}, finding that the sampler converges to the desired measure regardless of the particular hierarchy that is chosen.}
    \label{fig:hierarchyInd}
\end{figure}

\subsection{Altering Compactness}

\begin{figure}[h!]
    \begin{subfigure}{0.45\textwidth}
        \includegraphics*[width=\textwidth]{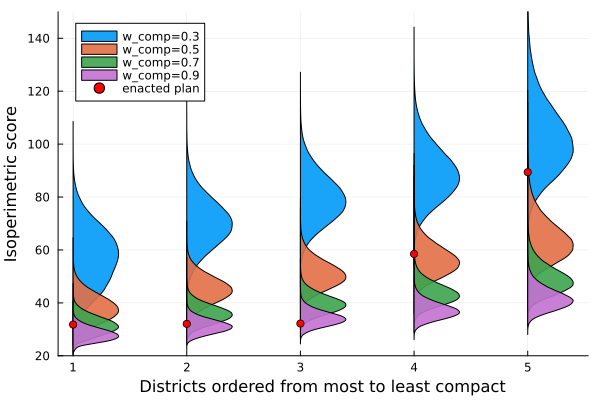}
        \caption{The typical isoperimetric ratios change as we alter the compactness weight $w_{comp}$. We also compare our results with the enacted plan.}
        \label{sfig:varyIsoWeightIsos}
    \end{subfigure}
    \hfill 
    \begin{subfigure}{0.45\textwidth}
        \includegraphics*[width=\textwidth]{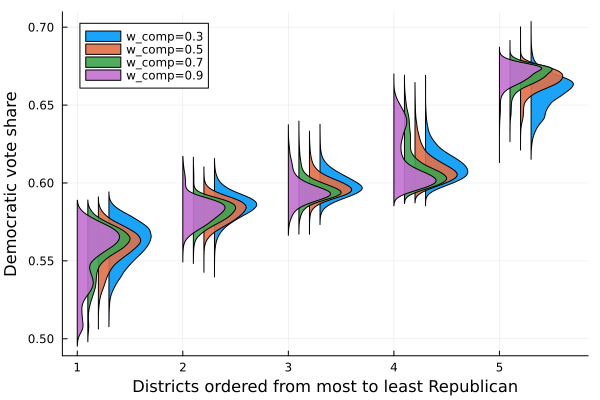}
        \caption{Changing the isoperimetric ratio affects the typical voting patterns across a range of compactness weights ($w_{comp}$).}
        \label{sfig:varyIsoWeightVotes}
    \end{subfigure}
    
    \begin{subfigure}{0.45\textwidth}
        \includegraphics*[width=\textwidth]{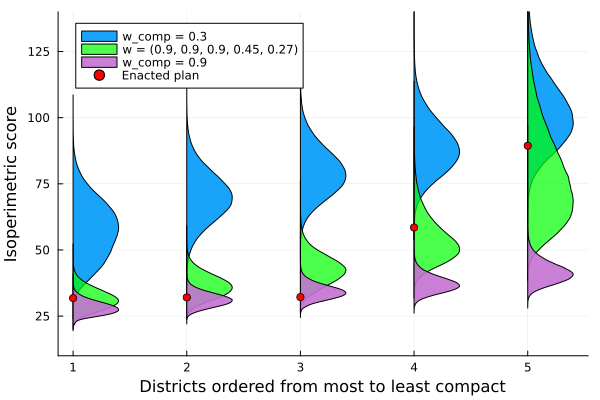}
        \caption{Using a weight vector for compactness, we are able to closely match the isoperimetric ratios of the districts in the enacted plan.}
        \label{sfig:tuneEnactedIso}
    \end{subfigure}
    \hfill 
    \begin{subfigure}{0.45\textwidth}
        \includegraphics*[width=\textwidth]{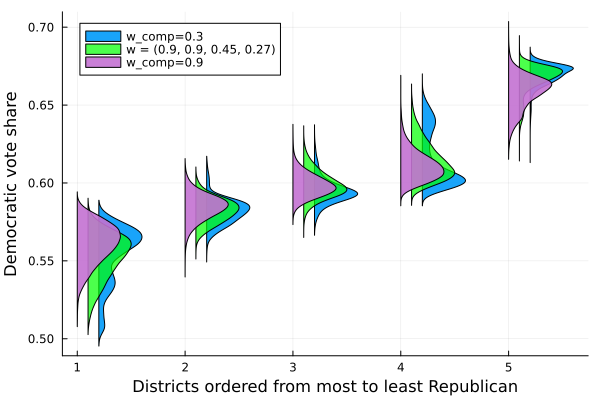}
        \caption{Using the weight vector results in a significantly different distribution over vote share outcomes than either the $w_{comp} = 0.3$ or 0.9 distributions.}
        \label{sfig:tuneEnactedIsoVotes}
    \end{subfigure}
    \label{fig:varyingiso}
    \caption{By varying the compactness component of the measure, we are able to control the marginal distributions on compactness in a fine-grained way, and observe the change in vote share distributions that arise as a result.}
\end{figure}

In this section, we investigate how that the choice of compactness weight and choice of compactness score can influence the typical vote shares. We also demonstrate how we can tune our compactness score to sample from different spaces of plans; in particular, we demonstrate that our sampling method provides the ability to control the shape of the rank-ordered compactness marginal distributions.

We begin by repeating our sampling procedure on three additional compactness weights for the target measure: We loosen the compactness consideration by examining $w_{comp}=0.3$ and strengthen the compactness consideration with two weights of $w_{comp}=0.7$ and $w_{comp}=0.9$. We repeat the convergence study for each of the different weights. We display the resulting typical levels of compactness in Figure~\ref{sfig:varyIsoWeightIsos}. We find that these changes lead to dramatic changes in the observed isoperimetric distributions (see Fig.~\ref{sfig:varyIsoWeightIsos}.  We also find that relaxing the compactness distribution can have significant effects on the typical voting patterns (see Fig.~\ref{sfig:varyIsoWeightVotes}).  For example, the three more compact ensembles almost never see a district with less than a 51\% Democratic vote share, whereas many of these districts are observed in the least compact distribution.

In Figure~\ref{sfig:tuneEnactedIso}, we compare the enacted congressional plan in Connecticut with our ensembles.  We see that there is a dramatic decrease in the two least compact districts. The three most compact districts agree well with the ensemble generated with $w_{comp}=0.9$ whereas the least compact district is more in-line with the resulting marginal distribution found with $w_{comp}=0.3$. There are many possible models that could potentially capture and/or explain this behavior. Let $J_{comp}^{PP}(\xi^{(j)})$ be the isoperimetric ratio of a single district and sort the districts so that $j=1$ is the most compact plan and $j=d$ is the least compact plan; then we could recast our energy in any of the following ways
\begin{align}
J_{comp; w}^{PP} &= \sum_{i=1}^d w_i J_{comp}^{PP}(\xi^{(j)}),\\
J_{comp; \alpha}^{PP} &= \sum_{i=1}^d J_{comp}^{PP}(\xi^{(j)})^\alpha,\\
J_{comp; m}^{PP} &= \sum_{i=1}^{d-m} J_{comp}^{PP}(\xi^{(j)}),
\end{align}
where $w=(w_1,...,w_d)$ is a weight vector with $w_i\geq w_j$ for $i<j$, $\alpha\in(0,\infty)$ generalizes the one norm that we have taken in \eqref{eq:PP} and places more or less emphasis on the most/least compact plans, and $m\in\{0,...,d-1\}$ is a way to consider the compactness score in the first $d-m$ plans and then ignore the remaining scores.  There are a variety of reasons why any one of these modifications may or may not be reasonable. The first score may, for example, model the redistricting committee finding it important to only have a tight compactness score for the first few plans and then to relax their criteria; the third is a special case of the first in which the weights are all zero or one. 

We make no claims about which model is best suited for the problem, but instead use the first variant to demonstrate that we have control over the marginal distributions on compactness. Specifically, we find that the weight vector $w=(0.9,0.9,0.9,0.45,0.27)$ aligns each of the rank ordered marginals with the enacted plan (see Figure~\ref{sfig:tuneEnactedIso}). 
% We remark that we have tuned these weights to center the least two compact districts in the enacted plan to fall in the center of the marginal distributions, but that this is likely not the optimal way to tune these weights as one should expect variations in the ranked vector of any given plan away from the mean of the ranked distributions. 
We leave the question of model inference and parameter inference for other work but note that one typically pays a cost in model inference when adding parameters and the trade-offs hear are unclear. The main purpose of this investigation is to show that we have flexibility in our class of distributions to match observed behaviors.

\subsection{Comparison to Spanning Forest Measures} 
We conclude by contrasting the samples on our new measure with samples that were generated on the measure weighted by spanning forest count. The spanning forest measure aligns with algorithmic considerations  when employing tree-based recombination methods such as Metropolized Forest Recombination \cite{autry2023metropolized,autry2021metropolized}, Reversible Recombination \cite{cannon2022spanning}, or an application of Sequential Monte Carlo \cite{mccartan2020sequential}.
Some have argued that these measures should be adopted into policy due to their correlations with discrete boundary lengths (see, e.g. \cite{deford2019recombination}), however, (i) we note that this measure may come with additional and potentially unintended policy considerations \cite{jonasBlogPost} and (ii) we stress caution when using convenient algorithmic choices to inform policy post hoc. 

Independent of the forest measure's policy relevance, we are now in a position to compare the consequences of the two measures on a larger graph. We use the Metropolized Forest Recombination method and run 4 independent chains for 500,000 steps, allowing up to a 2\% population deviation away from the ideal district size. We compare the marginal distributions across these 4 chains to validate convergence. 

\begin{figure}[ht]
    \begin{subfigure}{0.45\textwidth}
        \includegraphics[width=\textwidth]{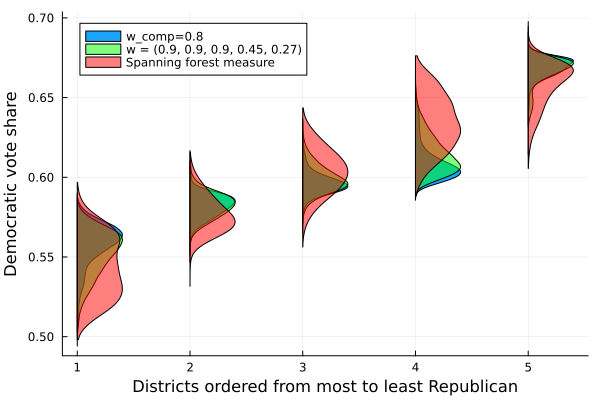}
        \caption{We compare the differences in the ordered marginal relative vote share distributions between the spanning forest measure and two of the measures we have investigated in this paper.}        
    \end{subfigure}
    \hfill 
    \begin{subfigure}{0.45\textwidth}
        \includegraphics[width=\textwidth]{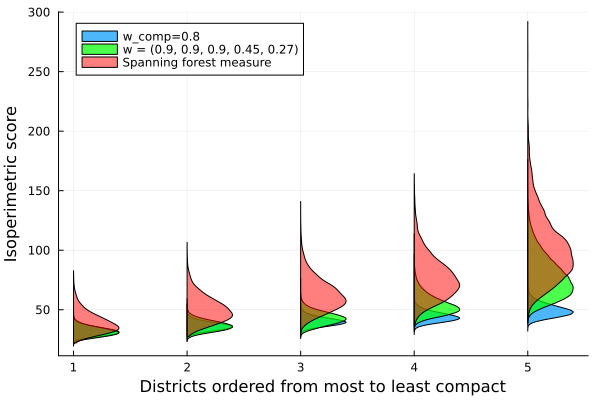}
        \caption{We compare the differences in the ordered marginal isoperimetric ratio distributions between the spanning forest measure and two of the measures we have investigated in this paper.}        
    \end{subfigure}
    \caption{The spanning forest measure differs significantly from the isoperimetric-based measure, in both vote share distribution and compactness.}
    \label{fig:compareSFtoNew}
\end{figure}

We compare the differences between the spanning forest measure and a selection of the measures above in Figure~\ref{fig:compareSFtoNew}. We find significant differences in the Polsby-Popper scores and the marginal vote fractions. In particular, under the spanning forest measure, one may see far more Democrats in the second most Democratic district and far fewer Democrats in the most Democratic district than compared with either of the presented distributions from our new measure. Similarly, one may expect far more Republicans in the two most Republican districts under the spanning forest measure than in the measures we have probed above. %These differences may lead to different conclusions when analyzing outliers.

\section{Discussion}

We have developed a novel multi-scale sampler that is capable of sampling redistricting plans on previously inaccessible policy-based probability measures. The sampler employs a single node flip algorithm at each scale and uses parallel tempering combined with a new swapping mechanism in order to exchange information across scales. The choice of sampling methodology is flexible. We re-iterate that the choice of hierarchy is arbitrary, however, the choice may affect the mixing times; we have developed several broad heuristics that can be used to generate a wide array of hierarchies. 

We demonstrate the efficacy of our sampler on the Connecticut congressional districts on a precinct level graph made up of roughly 700 nodes.  This is, by far, the largest graph that has been sampled when the measure on the partitions is not associated with spanning forests. Thus, our results demonstrate that our sampler mixes efficiently on a relevant class of measures at an unprecedented scale. The development of this method significantly expands our ability to both audit plans and compare policy implications in redistricting.

In the current work, we have focused on demonstrating the extension of our method to the traditional Polsby-Popper definition of compactness in redistricting. We have contrasted this with the recent concept of using cut-edges which have arisen due to their compatibility with forest-based measures. 
% In the case of recent Connecticut congressional districts, it seems like there is evidence that the cut-edge metric actually more consistent with the enacted plan than a Polsby-Popper score. 
In the future, we would also like to expand our method to sample from another recent way of measuring compactness which has shown to be aligned with human intuitive notions of compactness \cite{kaufman2021measure}.

In general, we believe that these methods will be able to be further extended to account for more traditional criteria such as the preservation of communities of interest and compliance with the Voting Rights Act. However, we also leave this extension for future work.

We remark that although there is flexibility in designing both the hierarchy and associated family of measures, there is also a sensitivity here on the mixing rate. We would like to continue to develop methods for designing hierarchies and associated measures to target fast mixing regimes in the extended state-space and additionally test the scaling of these methods to graphs of increasing size.

\section{Acknowledgements}
This material is based upon work supported by the National Science Foundation under Grant No. DMS-1928930 while the authors (GH and JCM) were in residence at the Mathematical Sciences Research Institute in Berkeley California, during the Fall 2023 semester. JCM thanks the Duke Analysis, Probability and Applied Mathematics RTG grant (DMS-2038056) for its suport. All authors thank an NC First grant for its suport. In particular, it partially supported  GH and GC during the 2022-2023 academic year.  
We are also thankful to the Rhodes Information Initiative at Duke (iiD) for its suport in hosting and supporting GC during his stay at Duke.

\bibliographystyle{plain}
\bibliography{hierarchy_swapping.bbl}

\appendix
\section{The Single Node Flip Operation}
\label{apdx:singlenodeflip}
Within each level of the hierarchy, we use a single node flip operation as the Metropolis-Hastings proposal. Single node flips, also called ``flip walks'', are used extensively in the literature and are closely related to Glauber dynamics (see \cite{MattinglyVaughn2014,Chikina_Frieze_Pegden_2017,chikina2019separating,fifield2020automated,herschlag2020quantifying,deford2019recombination,najt2021empirical}). 

A single node flip involves choosing an (node, district) pair from a set of \emph{valid flips}, where $(n, d)$ is a valid flip with respect to the current partition $\xi$ and population bounds $\delta$ if: 
\begin{itemize}
    \item $n$ is not a part of district $d$ i.e., $\xi(n) \neq d$.
    \item $n$ borders district $d$ i.e., there exists a neighbor $n'$ of $n$ where $\xi(n') = d$.
    \item Removing $n$ from district $\xi(n)$ would not cause $\xi(n)$ to become discontiguous i.e., $n$ is not an articulation point of district $\xi(n)$. 
    \item The districts resulting from assigning node $n$ to district $d$ are within population balance, i.e., $\text{pop}(d) + \text{pop}(n)$ and $\text{pop}(\xi(n)) - \text{pop}(n)$ fall in the range $\left( \frac{\text{total pop}}{\#\text{ districts}}(1 - \delta), \frac{\text{total pop}}{\#\text{ districts}}(1 + \delta)\right)$
\end{itemize}

The choice among valid flips may be uniform or tempered (that is, weighted in some way). In this work, we weight each flip is according to the energy of the districting that that flip would induce, exponentiated by 0.1. For example, at level $i$ in the hierarchy, we would propose a move $\xi_{i; j}$ with probability
\begin{align*}
\frac{\pi_{i}(\xi_{i; j})^{0.1}}{\sum_\ell \pi_{i}(\xi_{i; \ell})^{0.1}},
\end{align*}
where the sum in the demoniator ist taken over all \emph{valid flips}.
Note that, in order to Metropolize, the probability of choosing a given flip $\xi_1 \xrightarrow[]{(n,d)} \xi_2$ and the probability of choosing the reverse flip $\xi_2 \xrightarrow[]{(n,d')} \xi_1$ must \emph{both} be computed. 

When employing this algorithm, it is important to remove all articulation points from the graph.  For example, if a precinct is completely encircled by another precinct, then changing the district assignment of the encircling district would only work if the district was soley comprised of the two precincts. Since this is unlikely to happen, particularly with the population constraints, then the two precincts would be locked in their district assignment. To remedy this, we identify articulation points and merge any bi-connected components that are smaller than the size of a district.

\section{Details of Swapping Mechanism} 
\label{apdx:Swap}

Here, we briefly discuss relevant practical notes and details about the swap mechanism described in Section \ref{ssec:SwapMech}. Let $n$ be the number of steps made in each projection. 

\subsection{Coarse-to-Fine Projection}

On the coarse-to-fine partition $\xi_{C\to F}$, each step splits a non-split coarse node. At the $j$th step of the projection, we fix an allowable population deviation of $\delta_j = \delta_C + j \frac{d_F - d_C}{n}$. We choose probabilistically from the set of valid splits, where a split is a (node, district) pair. $(v, d) \in H_F \times [d]$ is a valid split if: 
\begin{enumerate}
    \item $v$'s parent has two children, $v$ and $v'$, and $y_\downarrow^{(j)}(v) = y_\downarrow^{(j)}(v') \neq d$ (that is, $v$ is currently assigned to the same district as its sibling), and 
    \item $(v, d)$ is a valid single node flip (see Appendix \ref{apdx:singlenodeflip}) with respect to $\xi_{C \to F}$ and population bounds $\delta_j$. 
\end{enumerate}

We weight the selection from among the set of valid splits according to the measure at the fine level exponentiated by $\alpha = 0.1$. That is, the probability that we choose a valid split is proportional to the weight of the resulting districting in the fine-level measure given by
\begin{align*}
\frac{\pi_{(i; j)}(y_\downarrow^{(j+1),k})^\alpha}{\sum_\ell \pi_{(i; j)}(y_\downarrow^{(j+1),\ell})^\alpha},
\end{align*}
where $y_\downarrow^{(j+1),k})$ are the possible refinement moves and $\pi_(i; j)$ is an intermediate measure with the compactness score $w^{(i)}_{comp}$ and population bound $\delta_\downarrow^{(k)}$.
 This biases the swap operation to make moves that have reasonable probability mass in the target (swapped) measure. An example is shown in Fig. \ref{fig:valid_splits}.

\subsection{Fine-to-Coarse Projection}

On the fine-to-coarse partition $\xi_{F\to C}$, each step merges a pair of split fine nodes. At the $j$th step of the projection, we fix an allowable population deviation of $\delta_F - j \frac{d_F - d_C}{n}$. We choose probabilistically from the set of valid merges, where a merge is characterized by a single coarse node. $v \in H_C$ is a valid merge if both of the following are true: 
\begin{itemize}
    \item $v$ has two children $c_1, c_2$ in the hierarchy, and
    \item $\xi_{F\to C}(c_1) = d_1 \neq \xi_{F\to C}(c_2) = d_2$ (that is, $c_1$ and $c_2$ are currently assigned to different districts). 
\end{itemize}
\emph{and} at least one of the following are true: 
\begin{itemize}
    \item $(c_2, d_1)$ is a valid single node flip with respect to $\xi_{F\to C}$ and population bounds $\delta_j$
    \item $(c_1, d_2)$ is a valid single node flip with respect to $\xi_{F\to C}$ and population bounds $\delta_j$
\end{itemize}

\begin{figure}[H]
    \centering
    \begin{subfigure}[T]{\textwidth}
        \centering
        \includegraphics*[height=100pt]{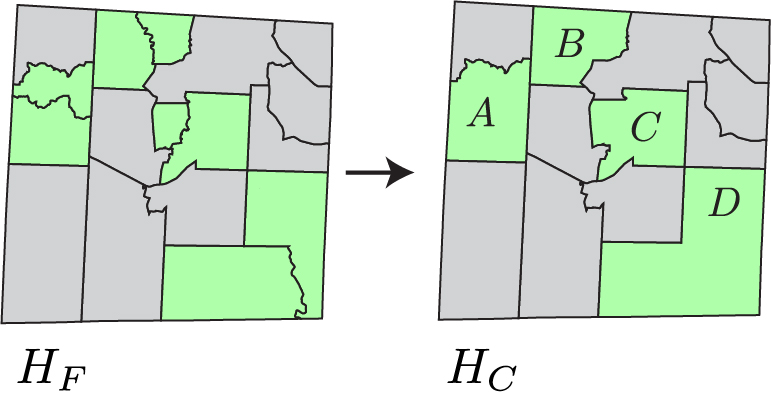}
        \caption{Fine and coarse hierarchy levels, with merged nodes highlighted in green.}
    \end{subfigure}
    \vspace{10pt}

    \begin{subfigure}[t]{0.45\textwidth}
        \centering
        \includegraphics*[height=100pt, clip=true, trim=0cm 0.4cm 0cm 0cm]{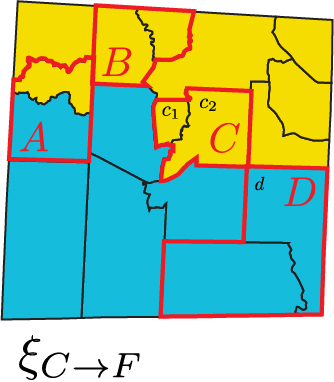}
        \caption{In this coarse-to-fine partition, the set of valid splits is $S = \{(c_1, \bluesq), (c_2, \bluesq), (d, \yellowsq)\}$. Node $A$ is already split; node $B$ cannot be split without disconnecting a partition. We choose a split from $S$, weighted according to the energy of the resulting partition.}
        \label{fig:valid_splits}
    \end{subfigure}
    \hfill
    \begin{subfigure}[t]{0.45\textwidth}
        \centering
        \includegraphics*[height=100pt, clip=true, trim=0cm 0.4cm 0cm 0cm]{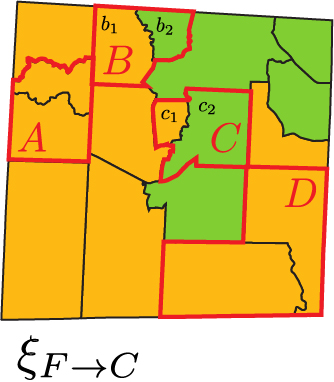}
        \caption{In this fine-to-coarse partition, the set of valid merges is $M = \{B, C\}$. Merging $B$ has two realizations: $(b_1, \greensq)$ and $(b_2, \orangesq)$. Merging $C$ has only one realization $(c_1, \greensq)$ because $(c_2, \orangesq)$ would disconnect the green partition. We choose a merge uniformly from $M$, and, if the result has multiple realizations, choose the lower-energy (e.g., more compact) one with probability $\frac23$.} 
        \label{fig:valid_merges}
    \end{subfigure}
    \caption{Examples of valid split and merge sets on partitions $\xi_{C\to F}$ and $\xi_{F\to C}$ on hierarchy levels $H_F, H_C$.}
\end{figure}

At each step, we choose \emph{uniformly} from among the set of valid merges. This is because, in the course of the fine-to-coarse projection, the set of merges that must be made is deterministic (every single split pair of nodes must be merged, and in fact this set determines $n$, the number of steps in both projections). However, it is possible that a given merge has two valid realizations (i.e., flipping $c_1$ to $d_2$ or $c_2$ to $d_1$); in such cases, we compare the energy of the resulting partitions and randomly choose the lower-energy (i.e., preferred) one with probability $\frac23$, and the higher-energy one otherwise. An example is shown in Fig. \ref{fig:valid_merges}.

\section{Hierarchy Generation Details}
\label{apdx:hierarchydetails}

We constructed two hierarchies on Connecticut: one deterministically-constructed 16-level hierarchy, generated according to the procedure in Section~\ref{sec:hierarchy}, and a 22-level randomized hierarchy following the principles in Section~\ref{sec:hierarchy}.1-\ref{sec:hierarchy}.4. 

%\subsection{Multipolygons and articulation points in the underlying graph}
%{\color{red} todo for greg? It's kinda addressed already (?)}

\subsection{Articulation Points}

\begin{figure}[H]
    \centering
    \includegraphics[width=0.45\textwidth]{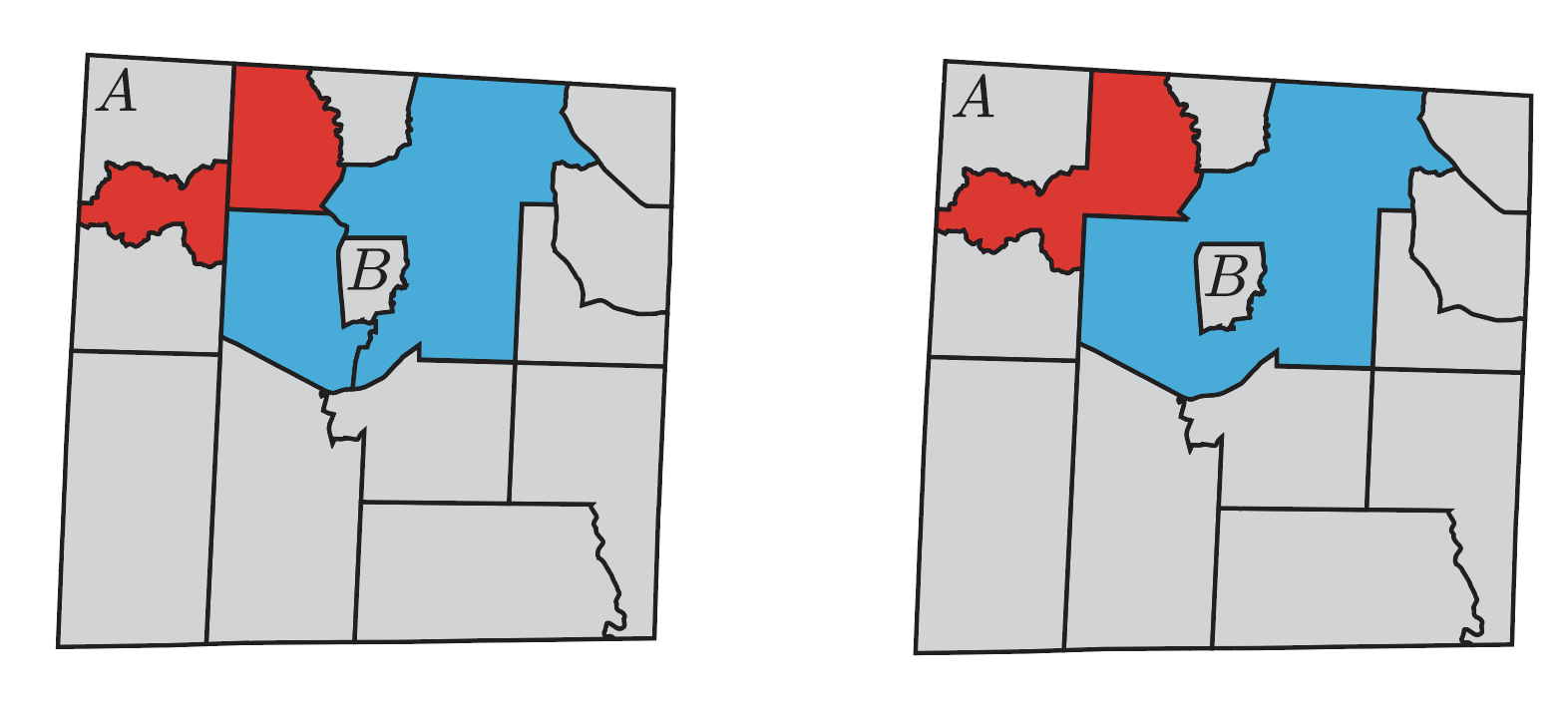}
    \caption{The pair of red nodes cannot be merged because doing so would create an articulation point separating node $A$ from the rest of the graph. Similarly, the pair of blue nodes cannot be merged, because doing so would create an articulation point separating node $B$ from the rest of the graph.}
    \label{fig:gen_noartpt}
\end{figure}

During construction of the hierarchy, we disallow merging any nodes that would form an articulation point in the resulting graph. Two examples of such disallowed merges are shown in Fig.~\ref{fig:gen_noartpt}.

\subsection{Regularized Population} 

\begin{figure}[H]
    \begin{subfigure}[h]{0.45\textwidth}
        \includegraphics*[width=\textwidth]{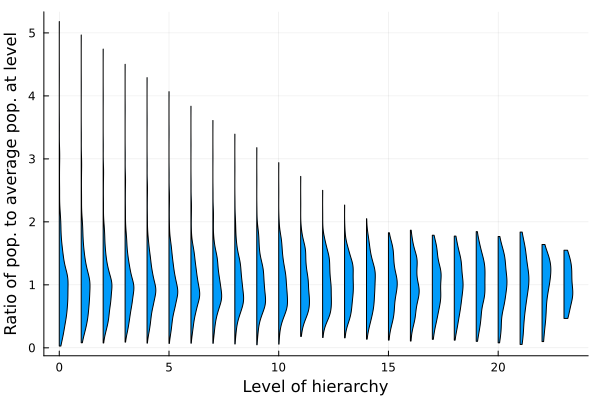}
        \caption{Distribution of node population at each level of the deterministically-constructed 24-level hierarchy on Connecticut, relative to the average node population at that level.}
    \end{subfigure}
    \hfill 
    \begin{subfigure}[h]{0.45\textwidth}
        \includegraphics*[width=\textwidth]{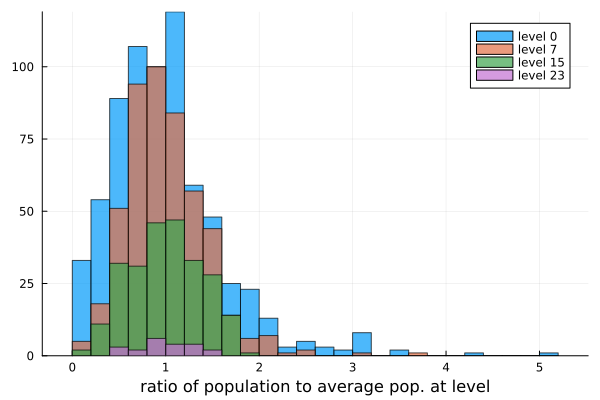}
        \caption{Distribution of node population at selected levels of the 24-level hierarchy, relative to the average node population at that level.}
    \end{subfigure}

    \begin{subfigure}[h]{0.45\textwidth}
        \includegraphics*[width=\textwidth]{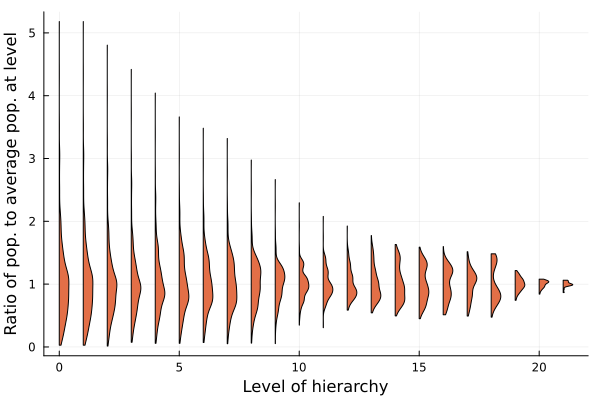}
        \caption{Distribution of node population at each level of the randomized 22-level hierarchy, relative to the average node population at that level.}
    \end{subfigure}
    \hfill 
    \begin{subfigure}[h]{0.45\textwidth}
        \includegraphics*[width=\textwidth]{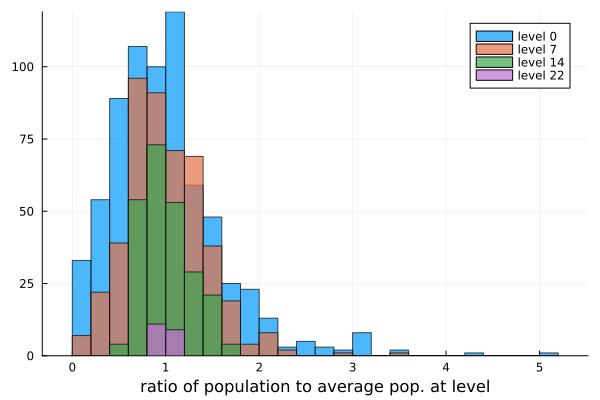}
        \caption{Distribution of node population at selected levels of the 22-level hierarchy, relative to the average node population at that level.}
    \end{subfigure}

    \caption{We construct tempering hierarchies such that, at higher levels of the hierarchy, the distribution of node populations becomes more tightly clustered around the average.}
    \label{fig:hierarchyPop}
\end{figure}

The gradually-regularized population of the 24-level and 22-level hierarchies on Connecticut are shown in Fig. \ref{fig:hierarchyPop}. As described in Section~\ref{sec:hierarchy}.2, we prioritize merging nodes whose combined population is near the average. 

\break
\subsection{Flat Compactness} 

\begin{figure}[H]
    \centering 
    \begin{subfigure}{0.45\textwidth}
        \includegraphics[width=\textwidth]{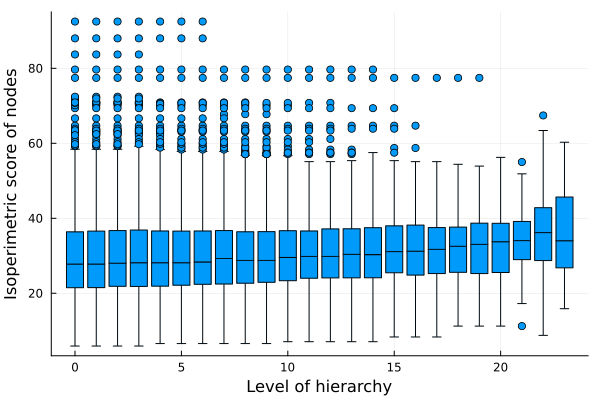}
        \caption{Isoperimetric scores of nodes at each level of the deterministically-constructed 24-level hierarchy on Connecticut. We seek to have each level have loosely similar distributions of compactness.}
    \end{subfigure}
    \hfill 
    \begin{subfigure}{0.45\textwidth}
        \includegraphics[width=\textwidth]{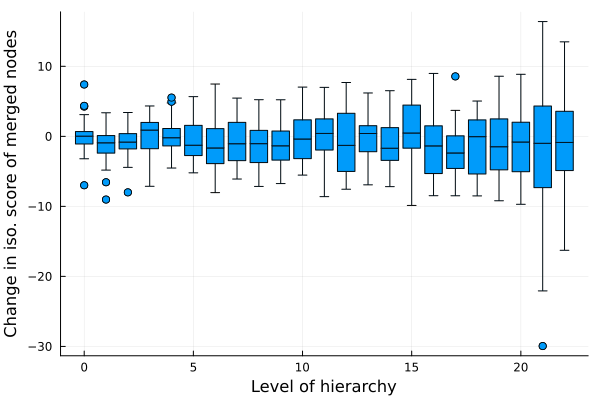}
        \caption{Change in compactness resulting from the merges at each level of the 24-level hierarchy.} 
    \end{subfigure}

    \begin{subfigure}{0.45\textwidth}
        \includegraphics[width=\textwidth]{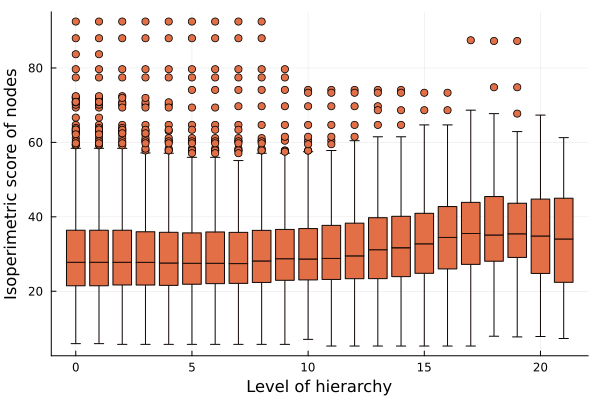}
        \caption{Isoperimetric scores of nodes at each level of the randomized 22-level hierarchy. We seek to have each level have loosely distributions ranges of compactness.}
    \end{subfigure}
    \hfill 
    \begin{subfigure}{0.45\textwidth}
        \includegraphics[width=\textwidth]{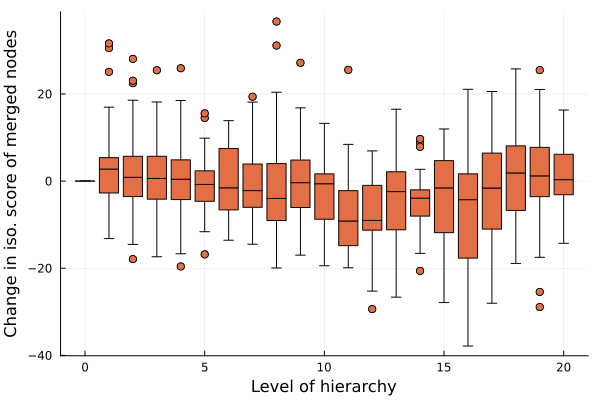}
        \caption{Change in compactness resulting from the merges at each level of the 22-level hierarchy.}
    \end{subfigure}
    \caption{We construct tempering hierarchies such that, especially at lower levels of the hierarchy, the compactness of the nodes is relatively stable. We do this by prioritizing merges whose resulting compactness is close to the compactness of the original (child) nodes.}
    \label{fig:hierarchyComp}
\end{figure}

The compactness of the nodes at each level of the 16-level and 22-level hierarchies on Connecticut are shown in Fig. \ref{fig:hierarchyComp}. As described in Section~\ref{sec:hierarchy}.4, we prefer to merge nodes where the merged compactness is similar to the average compactness of the original nodes, especially at the finest levels of the hierarchy (since the measures placed there have the largest compactness weight). We display the isoperimetric scores of the nodes at each level of each hierarchy in Fig.~\ref{fig:hierarchyComp}(a) and (c), and the difference between the compactness of the merged nodes at each level and the average compactness of their children (that is, the change in compactness that results from the merge) in Fig.~\ref{fig:hierarchyComp}(b) and (d).

\end{document}